\documentclass[12pt]{iopart}

\usepackage{iopams} 
\newtheorem{dfn}{Definition}
\newtheorem{prop}{Proposition}
\newtheorem{theorem}{Theorem}

\newtheorem{remark}{Remark}
\newtheorem{inference}{Inference} 
\begin{document}

\title[Bi-orthogonal spinors and time reversal in Clifford algebra]{Bi-orthogonal spinors associated to  $\mathcal{T}$-pseudo-Hermitian Rashba Hamiltonian : Time reversal in Clifford Algebra $Cl_3(\mathbb{R})$ and Pseudo-Supersymmetry}

\author{Arindam Chakraborty}

\address{Heritage Institute of Technology, Kolkata 700107, India}
\ead{arindam.chakraborty@heritageit.edu}
\vspace{10pt}
\begin{indented}
\item[]April 2023
\end{indented}

\begin{abstract}
A non-Hermitian version of Rashba Hamiltonian has been introduced motivated from the L\'evy-leblond type linearisation of Schr\"odinger equation in a Galilean invariant frame-work. The said Hamiltonian is found to be pseudo-Hermitian under fermionic time-reversal and  its eigen-spinors show bi-orthogonality. The whole discussion has been cast in a Clifford algebraic frame-work and the said bi-orthogonality has been variously understood in terms of time-reversal operator and Clifford involutions. The possibility of an equivalent version of Kramers' degeneracy theorem has been explored and the supersymmetric and pseudo-supersymmetric structures relating to the present system are also discussed.
\end{abstract}

%
%
%
%
%

\section{Introduction}
Exploring physical systems through quantum formalism involving non-Hermitian operators has  very much been in vogue for more than a couple of decades. In fact, the role of non-Hermitian Hamiltonians has got substantial attention in the study of complex potential, resonance phenomena in nuclear, atomic and molecular systems, systems with balanced gain and loss and even the systems which are not so quantum mechanical in sense but amenable to quantum methods (For example, classical statistical mechanical system, biological systems with diffusion, light propagation in wave guide)\cite{benderbook, moise11}. Nevertheless, non-Hermitian formalism has also engendered a phase of rethinking on traditional quantum methods \cite{baga15}. The demand of reality of the spectrum corresponding to the model Hamiltonian has put much emphasis on the aspect of space-time inversion symmetry ($\mathcal{PT}$-symmetry) in the works of several authors \cite{bender98, bender02, brody16, mosta02, mosta02a, mosta02b, znojil04}. In some typical examples the reality of the spectrum has been reported for multi-variable systems  admitting a variable specific notion of partial $\mathcal{PT}$ symmetry \cite{beygi15, chakraborty20, chakraborty21}. It has been revealed that a non-Hermitian operator $H$ in the conventional sense of inner-product can often show the so-called \textbf{pseudo-Hermiticity i.e.; $\eta H\eta^{-1}=H^{\dagger}$ for some $\eta$} and this fact is shown to have a close relation to the reality of the spectrum \cite{mosta02a, mosta02b}. The issue of bi-orthogonality of the eigenstates comes precisely in this context to set up a parallel framework  to deal with the pseudo-Hermitian Hamiltonians. Apart from developing formal methods \cite{brody14, curt07} bi-orthogonal formalism has been exploited even in real physical systems associated to super-symmetry, Lie super-algebra and quantum mechanics over Galois field etc. \cite{curt07a, houn17, oscar18, chang13}. 

The present article deals with a Hamiltonian involving a specific type of \textbf{Spin-Orbit Interaction (SOI)} known as \textbf{Rashba interaction} which is already well-studied in the contexts of semiconductor spintronics \cite{Ra60}, semiconductor hetero-structures \cite{Lu10}, superconductivity \cite{YG14},  topological quantum computation \cite{masa09}, transport properties of one dimensional quantum network \cite{zhang05}, quantum well \cite{zhang08, rashba88}, Kondo effect \cite{xu08} etc. Initially,  Rashba or \textbf{Bychkov-Rashba model} \cite{byrashba84} assuming a spin-orbit interaction $\mathbf{\sigma}\times\mathbf{P}\cdot\mathbf{u_3}$ ($\mathbf{u_3}$ being the unit vector along the $Z$-axis) was introduced in the contexts of electron resonances in 3-D semiconductors and momentum dependent splitting of spin bands in bulk crystals. Here, $\mathbf{\sigma}$ and $\mathbf{P}$ represent the Pauli spin matrices $\mathbf{\sigma}=\{\sigma_1, \sigma_2, \sigma_3\}$ and momentum operator $\mathbf{P}=\{-i\partial_1, -i\partial_2, -i\partial_3\}$ respectively. Recently, this model has gained attention with the emergence of nano-science where the study of quantum dots and rings \cite{kozin18} is of prime importance.  Our goal is to propose an extension of this  model  in the non-Hermitian regime considering the Hamiltonian to be \textbf{$\mathcal{T}$-pseudo-Hermitian} (i. e.; $\mathcal{T}^{-1}H\mathcal{T}=H^{\dagger}$), where $\mathcal{T}$ represents the fermionic time reversal operator \cite{jones10, bender11, jones14, choutri14}. The Hamiltonian so obtained is \textbf{isospectral} to the original Rashba Hamiltonian. Isospectral Hamiltonians have earlier been studied in the context of rotational symmetry \cite{lahiri88}, $\mathcal{PT}$-symmetry \cite{sinha02}, coherent states for factorizable Hamiltonians \cite{sanjay96}, super-symmetry \cite{naru97, pursey86} and many other fields \cite{pursey86a, gomes02, sid17}.

The \textbf{fermionic time-reversal} operator acts as an anti-involution i. e.; $\mathcal{T}^2=-\mathbf{1}$ and it has been extensively discussed in \cite{geru18, jones10, bender11, jones14, choutri14}. One of the important features of such time reversal operator is the \textbf{Kramers degeneracy theorem} \cite{klein51, scharf87, kruthoff19, rosch83, konstantin20, revaz09, chen22, lieu22} which stems from the existence of simultaneous eigenstates of the Hamiltonian and the time reversal operator. In the present scenario we observe that the time reversal operator does not commute with the non-Hermitian Hamiltonian thus prohibiting the possibility of Kramers' degeneracy.  However, a modified version of the said theorem has been obtained at least for the present case. The time reversal operator is of particular interest in the present occasion especially in the context of its close association with \textbf{involutions in Clifford algebra} $Cl_3(\mathbb{R})$ and inner-product associated with different automorphism groups. 

Motivated from the factorizability of the present Hamiltonian a \textbf{super-symmetric} formulation \cite{ junker19} is shown to be possible where the so called super-symmetric partners are different in the signs of the spin-orbit interaction terms. The study of super-symmetry has been extended to non-Hermitian domain by various authors \cite{supermosta01, supermosta02, alex20, superfring20, andri07, superznojil02}. The supercharges involve the generalised Clifford momenta. The super symmetric partner of present model involves repulsive interaction and has been used to study superconductivity \cite{wang17}.  It is interesting to note that two different Clifford momenta introduced here are pseudo-self-adjoint under fermionic time reversal. This paves the way to construct pseudo-super-symmetry as envisaged by Mostafazadeh \cite{supermosta01, supermosta02} and many other \cite{yes11, sinha08, bagchi15}. The related supercharges are found to be pseudo-self-adjoint under the action of \textbf{super time reversal operator}.  

The article has been organized in the following way : \textbf{Section-\ref{sectionbiortho}} describes the method of obtaining bi-orthogonal system in $\mathbf{C}^2$ and subsequent construction of bases of Clifford algebra ($Cl_3(\mathbb{R})$). \textbf{Section-\ref{sectionlevy}} elaborates the so-called \textbf{L\'evy-Leblond linearisation} scheme \cite{LL67, Hl99} which has been introduced to justify the existence of spin in the Galilean invariant framework (contra to Dirac theory of special relativistic electron) and hence its connection with non-relativistic Schr\"odinger equation \cite{monti06, nieder09, hue12, hue13}. This paves the way for an understanding of the present Hamiltonian in the framework of Clifford algebra \cite{Lo01, vaz16} as has been discussed by the present author elsewhere \cite{chakraborty22} for many other Hamiltonians. A generalised form of non-Hermitian Rashba Hamiltonian has been obtained in \textbf{section-\ref{cliffmomentum}} starting from a linearised version of Sch\"odinger equation involving generalised Clifford momentum and admitting four-component spinor solutions. The present Hamiltonian is obtained for some suitable choice of constants. \textbf{Section-\ref{biortho}} discusses the spectrum and eigen-spinors of the Hamiltonian and bi-orthogonality of the said spinors.  In the present case the eigen-spinors of the Hamiltonian and its adjoint form a \textbf{two-parameter bi-orthogonal set} unlike the set used to construct $Cl_3(\mathbb{R})$ generators in \textbf{Section-\ref{sectionbiortho}}. The associated bi-orthogonal spinor projection operators are also constructed. The issues of pseudo-Hermiticity and time reversal have been dealt with in \textbf{section-\ref{sectionpseudo}} which also explores the possibility  of different types of bi-orthogonality in Clifford algebraic frame-work and their relation with time reversal. In this section a spinor has been identified as left ideal of $Cl_3(\mathbb{R})$ algebra and spinor operators have been understood in terms of direct and time reversed generators of even sub-algebra  $Cl_3^+(\mathbb{R})$.  A  possible analogue of Krammers theorem is also conjectured in this section.  Finally, the notions of \textbf{super-symmetry (SUSY)} and \textbf{pseudo-super-symmetry (pseudo-SUSY)} have been elaborated in \textbf{section-\ref{sectionsusy}}. 

Apart from emphasising spin  in a typical Galilean invariant framework  the choice of Rashba interaction as a viable model of exploring aspects of pseudo-Hermiticity, bi-orthogonality and time-reversal can be justified through the following four grounds:

I. The present construction considers the Rashba Coefficient (RC)  a free parameter unlike the construction motivated from special relativity. This speaks in favour of greater experimental viability where  it is not far less customary to study a system for different values of RC.

II. The Hamiltonian involves tensor product operators involving spin and linear momentum, where the former is responsible for non-Hermiticity and the later is Hermitian for obvious reason leading to $\mathcal{T}$-pseudo-Hermitian interaction term. The eigenvectors of momentum operator, though not square integrable,  provide a complete set of bases. This has been explained following Gel'fand \cite{gelfand64} in view of an analogous theorem in finite dimension and Fourier theory.   (see section-\ref{biortho}).  

III.  It helps us to explore the role of time reversal in a fermionic system and Kramers like results in pseudo-Hermitian Quantum Mechanics which are two staples in the study of its Hermitian counterpart.  Furthermore, the effect of time reversal is intrinsically associated with complex conjugation which, by definition, is an involution in $\mathbf{C}$ as an algebra of complex number. The use of \textbf{Clifford involutions} widens the scope of involution to express the role of time reversal when the spinor wave function is viewed as an element of the left ideal of $Cl_3(\mathbb{R})$.

IV. The presence of pseudo-Hermiticity, though does not change the energy spectrum as compared with its Hermitian counterpart,  greatly modifies the expressions of the spinor wave functions associated with it and hence influences the alignments of spin. The  Rashba interaction term has already been understood as something  introducing spin precession and related phenomena \cite{wang04, aver02, long16}. The present model  is worthy of future investigation both for theoretical and experimental contexts, where, especially in the present day scenario spin injection and various other spin related manipulations are preferred over the conventional possibilities (like changing the electric field etc.) available in the field of spintronics.

\section{Auerbach Bi-orthogonal System and  $Cl_{3}(\mathbb{R})$ generators}\label{sectionbiortho}

\begin{dfn}
	Two pairs of vectors $\{\vert\phi_j\rangle : j=1,2\}$ and $\{\vert\chi_j\rangle : j=1,2\}\in\mathbf{C^2}$ are said to be bi-orthogonal if $\langle\phi_j\vert\chi_k\rangle=\delta_{jk}$.
\end{dfn}

The condition of bi-orthoganality in $\mathbf{C}^2$ can be given by the following theorem

\begin{theorem}
	Given any pair of vectors  
	$\{\vert v_j\rangle = 
	\left(\begin{array}{c}
		c_j^{(1)} \\
		c_j^{(2)}
	\end{array} \right): j = 1,2\}\in \mathbf{C^2}$ two pairs of vectors $\vert\phi_j\rangle=T\vert v_j\rangle$ and $\vert\chi_j\rangle=(T^{-1})^{\dagger}\vert v_j\rangle$ constitute a bi-orthogonal system under the action of a transformation $T = (\cos \frac{\theta}{2} ){\bf 1}_2 +
	2(\cos \frac{\phi}{2}  \sin \frac{\theta}{2} )\sigma_1 - 2(\sin \frac{\phi}{2}  \sin \frac{\theta}{2} )\sigma_2$  provided $\langle v_j\vert v_k\rangle=(c_j^{(1)})^{\star}c_k^{(1)}+(c_j^{(2)})^{\star}c_k^{(2)}=0\:\:\forall\:\: j\neq k$.
\end{theorem}

\textbf{Proof} : Since $T=T^{\dagger}$ with the present sense of inner-product $\langle\phi_j\vert\chi_k\rangle=\langle v_j\vert T(T^{-1})^{\dagger}\vert v_k\rangle=\langle v_j\vert T(T^{\dagger})^{-1}\vert v_k\rangle=\langle v_j\vert v_k\rangle$. Hence the theorem follows from the notion of bi-orthogonal system stated above. $\square$
	
\vspace{0.5cm}	

\begin{remark}
	According to Auerbach such a bi-orthogonal system is always available in any finite dimensional vector space \cite{hajek08}.
\end{remark}

Now, taking , $\{\vert v_j\rangle = 
\frac{1}{\sqrt{2}}\left(\begin{array}{c}
	1 \\
	(-1)^{j-1}
\end{array} \right): j = 1,2\}\in \mathbf{C^2}$, $\phi=\pi$ and $\gamma=\sqrt{1-\omega^2}=\sin\theta$ we get 

\begin{equation}\label{pgamma}
	\sigma_m^{\gamma} = {i^{m+1}}\sum_{j, k=1}^2{c_{jk}^{(m)}}\vert\phi_j\rangle \langle\chi_k \vert : m = 1, 2, 3,
\end{equation}

where, $c_{jk}^{(1)} = (-1)^j \delta_{jk}$ and $c_{jk}^{(3)} = (-1)^j c_{jk}^{(2)} = (1 -(-1)^j c_{jk}^{(1)})$. These matrices are Hermitian in the sense of conventional inner-product $\langle v_j\vert v_k\rangle=(c_j^{(1)})^{\star}c_k^{(1)}+(c_j^{(2)})^{\star}c_k^{(2)}$ defined on $\mathbf{C^2}$.

\begin{dfn}
	The Clifford algebra $Cl_{3}(\mathbf{\mathbb{R}})$ is a real associative algebra equipped with the operation $ab=a\cdot b+a\wedge b$ and admitting the set of generators $\mathcal{G}=\{{1},e_1, e_2, e_3, e_{12}, e_{23}, e_{31}, e_{123}\}$ satisfying the following relations:
	\begin{eqnarray}
		e_i^2=1\:\:\forall\:\: i=1,2, 3.\:\:{\rm{and}}\:\:
		e_{ij}=e_ie_j=-e_je_i=-e_{ji}\:\:\forall\:\: i\neq j.
	\end{eqnarray}
\end{dfn}

Identifying the set of Pauli spin matrices $\{\sigma_m : m=1, 2, 3\}=\{e_m : m=1, 2, 3\}$, The generators for $Cl_{3}(\mathbb{R})$ can be written as $\mathcal{G}^0=\{\mathbf{1}_2,\sigma_1, \sigma_2, \sigma_3, i\sigma_3, i\sigma_1, i\sigma_2, i\mathbf{1}_2\}$ and the isomorphism $Cl_3(\mathbb{R})\cong \mathbf{R}\bigoplus \mathbf{R^3}\bigoplus\bigwedge^2\mathbf{R^3}\bigoplus\bigwedge^3\mathbf{R^3}$ is possible with the exterior algebra $\bigwedge \mathbf{R^3}$ \cite{Lo01}.

Now, writing explicitly we get from equation-\ref{pgamma}

\begin{equation}
	\sigma_1^{\gamma}=\omega^{-1}\left(\begin{array}{cc}
		-i\gamma & 1 \\
		1 & i\gamma
	\end{array} \right), \:\:\:\sigma_2^{\gamma}=\left(\begin{array}{cc}
	0 & -i \\
	i & 0
\end{array} \right) {\rm{and}} \:\:\:\sigma_3^{\gamma}=\omega^{-1}\left(\begin{array}{cc}
1 & i\gamma \\
i\gamma & -1
\end{array} \right).
\end{equation}
 It is obvious to note that $\{\sigma_1^{\gamma}, \sigma_3^{\gamma}\}$ are non-Hermitian in the conventional sense of inner-product. In view of the above discussion a new set of $Cl_3(\mathbb{R})$ generators $\mathcal{G}^{\gamma}=\{\mathbf{1}_2, e_1^{\gamma}, e_2^{\gamma}, e_3^{\gamma}, e_{12}^{\gamma}, e_{23}^{\gamma}, e_{31}^{\gamma}, e_{123}^{\gamma} \}$ can be formed. Such generators resemble the previous one for $\gamma=0$. The algebra $Cl_3$ has the decomposition $Cl_3=Cl_3^+\bigoplus Cl_3^-$
 called even and odd parts respectively. $Cl_3^+$ is non-commutative and isomorphic to quaternion algebra $\mathbb{H}$. The generators of $Cl_3^+$ is given by $\mathcal{G}_+^{\gamma}=\{\mathbf{1}_2, e_{12}^{\gamma}, e_{23}^{\gamma}, e_{31}^{\gamma}\}$ and it is of particular interest in the present context as discussed in the following section.
 
\section{ L\'evy-Leblond linearisation, Galilei Group and Clifford algebra} \label{sectionlevy}
Let us consider the time dependent Schr\"odinger equation for a single free particle in three dimension given by
\begin{equation}\label{wave1}
	\frac{\mathbf{P}^2}{2}\psi=i\partial_t\psi\:\:\ {\rm{where}},\:\:i=\sqrt{-1}. 
\end{equation} 
 
Here, $\mathbf{P}^2 \equiv -\nabla^2$ with $\mathbf{P}=(P_1, P_2, P_3)=(-i\partial_{x_1}\:\:-i\partial_{x_2}\:\:-i\partial_{x_3})$ and $\psi\dot{=}\psi(x_1, x_2, x_3, t)$, the wave function is a complex valued function of spatial coordinates $(x_j\vert j=1, 2, 3)$ and time $t$. The invariance of localized probability density i. e. ; $\vert\psi^{\prime}(\vec{x}^{\prime}, t^{\prime})\vert^2=\vert\psi(\vec{x}, t)\vert^2$ ensures the invariance of the equation under the action of Galilei group as defined below.

\begin{dfn}
The Galilei group $G(1, 3)$ is defined by the space $(x_1, x_2, x_3)\in\mathbf{R}^3$ and time $t\in\mathbb{R}$ transformation 
\begin{eqnarray}
	x_j^{\prime}=R_{jk}x_k+v_jt+a_j\:\:{\rm{and}}\:\:
	t^{\prime}=t+b : j=1, 2, 3.
\end{eqnarray}
Here, $a_j$, $b$ and $v_j$ are real parameters of space translation, time translation and pure Galilei transformation respectively with the matrix $(R_{jk})$ specifying rotation. 	
\end{dfn}

\begin{remark}
The case for $a_j=b=0$ corresponds to the homogeneous Galilei group $HG(1, 3)$. The group $G(1, 3)$ is the semi-direct product of its Abelian group generated by space and time translation with $HG(1, 3)$\cite{ monti06, nieder09, hue12, hue13}.	
\end{remark}

\begin{theorem}
	Under the unitary ray representation of the Galilei group the two wave functions $\psi$ and $\psi^{\prime}$ are related by
\begin{equation}
\psi^{\prime}(\vec{x}^{\prime}, t^{\prime})=e^{if(x, t)}\psi(\vec{x}, t),
\end{equation}
where, $f(\vec{x}, t)=\frac{1}{2}v^2t+v_jR_{jk}x_k+c$.
\end{theorem}

\textbf{Proof} : 
	It is obvious to check that 
	\begin{eqnarray}
		i\partial_{t^{\prime}}\psi^{\prime}&=&e^{if}[i\partial_t\psi-iR_{kj}v_k\partial_{x_j}\psi+(R_{kj}v_k\partial_{x_j}f-\partial_tf)\psi]\nonumber\\
		\partial_{x_j^{\prime}}\psi^{\prime}&=&R_{jk}e^{if}(i\partial_{x_k}f\psi+\partial_{x_j}\psi).
	\end{eqnarray}
Hence the expression of $f(\vec{x}, t)$ in view of equation-\ref{wave1} $\square$.

 \vspace{0.5cm}
 
It has been argued elsewhere that the attempt to linearise a Galilean invariant equation leads to four-component spinor solution of the Schr\"odinger equation 
\begin{equation}\label{wavespinor}
	\frac{1}{2} \mathbf{P}^2 \mathbf{1}_4{\bf \Psi}=E\bf{\Psi} 
	\label{a1}
\end{equation} 
 and this automatically gives rise to the demand of a Clifford algebraic frame-work as revealed by the following theorem due to L\'evy-Leblond.
 
 \begin{theorem}
 Let, $\bf \Psi\dot{=}(\psi\:\:\eta)^{\dagger}\dot{=}(\psi_1\;\:\psi_2\:\:\eta_1\:\:\eta_2)^{\dagger}$ be a four-component spinor solution of equation-\ref{wavespinor}. The coupled equations satisfied by $(\psi\:\:\eta)^{\dagger}$ are given by
\begin{eqnarray}\label{wavespinor0}
 		{\rm{(a)}} \:\:e_jP_j\psi+2i\eta=0 \:\:\:{\rm{(b)}}.\:\: e_jP_j\eta-iE\psi=0.                  
 \end{eqnarray}
 \end{theorem}
\textbf{Proof} : 
	Considering $\{L^{\prime}, L; M_j^{\prime}, M_j; N^{\prime}, N\}$ as $4\times 4$ matrices we write equation-\ref{wavespinor} as
	\begin{eqnarray}
		[(L^{\prime}E+M_i^{\prime}P_i+N^{\prime})(LE+M_jP_j+N)-(2E-P_kP_k)]\bf{\Psi}=0.
	\end{eqnarray}
 
Equating coefficients of $E$ and $P_j$'s in Eq.(\ref{wavespinor}) we obtain
\begin{eqnarray}
	L^{\prime}L=0,  ~~~~  N^{\prime}N=0,~~~~L^{\prime}N+N^{\prime}L=2, ~~~~    L^{\prime}M_j+M_j^{\prime}L=0, ~~ \nonumber  \\
	M_i^{\prime}M_j+M_j^{\prime}M_i=-2\delta_{ij},   ~~~~~~N^{\prime}M_i+M_i^{\prime}N=0.
	~~~~~~ (i, j=1,2,3) ~~
	\label{a3}
\end{eqnarray} 
Identifying $M_4=i(L+\frac{1}{2}N)$, $M_4^{\prime}=i(L^{\prime}+\frac{1}{2}N^{\prime})$, $M_5=L-\frac{1}{2}N$, $M_5^{\prime}=L^{\prime}-\frac{1}{2}N^{\prime}$,
the relations (\ref{a3}) can be represented in a condensed form
\begin{equation}\label{wavespinor1}
	M_i^{\prime}M_j+M_{j}^{\prime}M_i=-2\delta_{ij} ~~~~~~\forall~ i, j=1,...,5.
\end{equation} 

Choosing an invertible matrix $\Lambda$ such that $M_5=-i\Lambda$, $M_5^{\prime}=-i\Lambda^{-1}$, $M_j=\Lambda\gamma_j$ and $M_j^{\prime}=-\gamma_j\Lambda^{-1}, ~~\forall~j=1,...,4$,  consistent with Eq.(\ref{wavespinor1}), where the $\gamma$-matrices are defined as 
 
\begin{eqnarray}\label{gamma}
	\gamma_{j}= \left(\begin{array}{cc}
		0 & e_j    \\
		e_j & 0
	\end{array} \right)
	~~~ \rm{and} ~~
	\gamma_4=\rm{diag}(1,1,-1,-1).  \nonumber
\end{eqnarray} 
 \begin{eqnarray}
 	\Lambda= \left(\begin{array}{cc}
 		0 & \mathbf{1}_2 \\
 		\mathbf{1}_2 & 0
 	\end{array} \right)   \nonumber
 \end{eqnarray}
 and hence calculating the $L, M, N$ matrices, we get equation-\ref{wavespinor0}. $\square$
 
 \vspace{0.5cm}

\begin{remark}
For our present purpose, it is to be noted that no loss of generality would happen upto this point of discussion if the set $\{e_j\vert j=1, 2,3\}$ is replaced by the set $\{e_j^{\gamma}\vert j=1, 2, 3\}$. However, this replacement will introduce certain physical and mathematical meanings to look at the quantum world in a characteristically different way. 
\end{remark} 
 
\section{Generalized Clifford Momentum and Non-Hermitian Rashba Hamiltonian}\label{cliffmomentum}

Let us begin with the following definition.
\begin{dfn}
An operator valued $1$-blade of the form $\wp_{\gamma}^Q=e^{\gamma}_j(P_j+Q_j)$, where $P_j=-i\partial_{x_j}$ is the $j$-th component of linear momentum and $Q_j$ being any function of $\{x_k : k=1,2,3\}$ can be called a generalized \textbf{Clifford momentum} {\footnote{Clifford momentum is defined in the spirit that spin-orbit interactions in many cases including the  present one is equivalent to expressing the generalized  momentum in Clifford Bases. This fact also reasserts and generalizes Hertz's contention of  kinetic origin of the potential \cite{deri10, lanczos70} even in case of non-classical interactions. }} expressed in terms of Clifford generators $\{e_j^{\gamma} : j=1, 2, 3\}$. 
\end{dfn}

\begin{theorem}
	Given the linearized equations:
	\begin{eqnarray}\label{bi}
	{\rm{(a)}}\:\:\:	\wp_{\gamma}^A\psi+2i\eta=0, \:\:\:    	{\rm{(b)}}\:\:\wp_{\gamma}^B\eta-iE\psi=0
\end{eqnarray}
	with $\psi=(\psi_1\:\:\:\psi_2)^{\dagger}$ and $\eta=(\eta_1\:\:\:\eta_2)^{\dagger}$
	and an operator-valued Clifford product acting on a two component function
$\Phi$,  $\wp_{\gamma}^A\wp_{\gamma}^B\Phi=\wp_{\gamma}^A\rfloor\wp_{\gamma}^B\Phi+\wp_{\gamma}^A\wedge\wp_{\gamma}^B\Phi$ with $'\rfloor'$ and $'\wedge'$ as  contraction and exterior product respectively, the $4\times 4$ matrix  Schr\"odinger equation is given by
\begin{eqnarray}\label{matrix}
	H_{\gamma}{\bf \Psi}= \left(\begin{array}{cc}
		H_{\gamma}^{AB} & 0 \\
		0 & H_{\gamma}^{BA}
	\end{array} \right){\bf \Psi}=E{\bf \Psi}.
\end{eqnarray}
Here, $H_{\gamma}^{AB}=\frac{1}{2}\wp_{\gamma}^B\wp_{\gamma}^A$ and $H_{\gamma}^{BA}=\frac{1}{2}\wp_{\gamma}^A\wp_{\gamma}^B$ and $\Psi=(\psi\:\:\eta)^{\dagger}$.
\end{theorem}

\textbf{Proof} : Actions of $\wp^B_\gamma$ and $\wp^A_\gamma$ from the left on equation-\ref{bi} (a) and (b) respectively lead to two Schr\"odinger equations
\begin{eqnarray}
	H_{\gamma}^{AB}\psi=E\psi          \nonumber  \\
	H_{\gamma}^{BA}\eta=E\eta,
	\label{b2}
\end{eqnarray}
or simply

\begin{equation}
	H_{\gamma}{\mathbf{\Psi}}=E{\mathbf{\Psi}}.
\end{equation}

Explicitly, Eq.(\ref{b2}) may be written as
\begin{equation}
	\frac{1}{2}[(e_j^{\gamma})^2\{P_j^2+(P_jA_j+B_jP_j+B_jA_j)\}+e^{\gamma}_{jk}(P_jA_k+B_jP_k+B_jA_k)]\psi=E\psi,
	\label{b3}
\end{equation}
\begin{equation}
	\frac{1}{2}[(e_j^{\gamma})^2\{P_j^2+(P_jB_j+A_jP_j+A_jB_j)\}+e^{\gamma}_{jk}(P_jB_k+A_jP_k+A_jB_k)]\eta=E\eta.
	\label{b4}
\end{equation}

\vspace{0.5cm}

Choosing $\mathbf{A}=\mathbf{B}^{\star},=-i(\alpha_1\:\:\alpha_2\:\:\alpha_3); \alpha_j\in\mathbb{R} : j=1, 2, 3$,  we get the generalized Rashba Hamiltonian
\begin{eqnarray}
H_{\gamma}^{AB}=\nonumber\\
\frac{1}{2}[P^2+\alpha^2]\mathbf{1}_2\nonumber\\
+i[e^{\gamma}_{12}(\alpha_1P_2-\alpha_2P_1)+e^{\gamma}_{23}(\alpha_2P_3-\alpha_3P_2)+e^{\gamma}_{31}(\alpha_3P_1-\alpha_1P_3)].
\end{eqnarray}

We shall consider an analogue of conventional Rashba Hamiltonian (that can be recovered on letting $\gamma=0$) by choosing $P_3=0$ and $\alpha_1=0=\alpha_2, \alpha_3=\beta$. This choice leads to

\begin{eqnarray}\label{rashba1}
	R^+_{\gamma}=H^{AB}_{\gamma}=\frac{1}{2}(P_1^2+P_2^2+\beta^2)\mathbf{1}_2+i\beta(e_{31}^{\gamma}P_1-e_{23}^{\gamma}P_2).
\end{eqnarray} 

The Hamiltonian $R^+_{\gamma}$ is not Hermitian since $(R^+_{\gamma})^{\dagger}=R^+_{-\gamma}$. Where,
\begin{eqnarray}\label{rashba2}
	R^+_{-\gamma}=H^{AB}_{-\gamma}=\frac{1}{2}(P_1^2+P_2^2+\beta^2)\mathbf{1}_2+i\beta(e_{31}^{-\gamma}P_1-e_{23}^{-\gamma}P_2).
\end{eqnarray} 

Replacing the \textbf{Rashba Coefficient (RC)} $\beta\rightarrow -\beta$ we obtained the corresponding cases with $R_{\gamma}^-=H_{\gamma}^{BA}$ and $R_{-\gamma}^-=H_{-\gamma}^{BA}$. The Hamiltonians $R_{\pm\gamma}^+$ reduce to the conventional Rashba Hamiltonian for $\gamma-0${\footnote{This method is not usually followed in conventional discussion on spin-orbit coupling (for example in \cite{winkler03}) where the present Hamiltonian has been shown as an approximation of more general expression of a special relativistic Hamiltonian. The present derivation seems to be more realistic since it imposes no restriction on Rashba Coefficient ($\beta$) providing better experimental viability. }}.

\begin{remark}
The RC comes into existence when an electric field is applied perpendicular to the plane of the 2D material, for example metallic graphene or Black Phosphorous (BP). The strength of the interaction is often manipulated by changing the said electric field. 
\end{remark}

\begin{remark}
Similar expression of Hamiltonian may be obtained for the same system in the presence of a magnetic field $\mathbf{B}=(0, 0, B_3)$ introducing Zeeman term \cite{wu05, vasil06}. Choosing $\wp^\pm=e_j^{\gamma}(P_j+A^v_j\pm i\alpha_j)$ the Hamiltonian for unit electric charge is given by

\begin{eqnarray}
H^\pm=\wp^\pm\wp^\mp &=& \frac{1}{2}[(P_1+A_1^v)^2+(P_2+A_2^v)^2] \mathbf{1}_2\nonumber\\
&\pm & i\beta[e_{31}^\gamma (P_1+A_1^v)-e_{23}^\gamma( P_2+A^v_2)]+e_{3}^\gamma B_3.
\end{eqnarray}

Here, $\mathbf{A}^v=(A^v_1, A^v_2, 0)$ being the vector potential and $\mathbf{B}=i\mathbf{P}\times\mathbf{A}^v$. The pseudo-Hermiticity of $H^\pm$ may be verified immediately considering $\mathcal{T}^{-1}\mathbf{B}\mathcal{T}=-\mathbf{B}$. (see section-\ref{sectionpseudo})
\end{remark}

 $R_0^{\pm}$ and $R^{\pm}_{\pm\gamma}$   are isospectral but they have different eigenvectors as discussed in the following section.

\section{Spectrum of $R_{\pm\gamma}$ and two-parameter bi-orthogonal eigenspinors}\label{biortho}
Considering the eigenvector of the form $\vert\psi\rangle\rangle=\vert A\rangle\otimes\vert\mathbf{p}) (\in \mathcal{H}^{\otimes}\dot{=}\mathbf{C^2}\otimes\mathcal{H}_0)$ for both $R^+_{\pm\gamma}$, where, $\vert A \rangle\in\mathbf{C}^2$ is a column vector and $\vert\mathbf{p})\in\mathcal{H}_0$ is the free-particle part. With de-Broglie relation $\vert\mathbf{p}\vert=\vert\mathbf{k}\vert$ we find the eigenvalues corresponding to the tensor product ansatz $\vert\psi(x)\rangle\rangle=
	 \frac{1}{2\pi}\left(\begin{array}{c}
		a_1  \\
		a_2 
	\end{array} \right)\otimes e^{i\mathbf{p}\cdot \mathbf{x}}$ in co-ordinate representation with $a_1, a_2\in\mathbb{C}$, $\mathbf{p}\cdot \mathbf{x}=p_1x_1+p_2x_2$ and $(\mathbf{x}\vert\mathbf{p})=\frac{1}{2\pi}e^{i\mathbf{x}\cdot\mathbf{p}}$. Let us call $\vert A\rangle=\left(\begin{array}{c}
		a_1  \\
		a_2 
	\end{array} \right)$, the \textbf{finite part} and $ e^{i\mathbf{p}\cdot \mathbf{x}}$, the \textbf{free-particle part} of a typical eigenspinor. The eigenvalues of $R^+_{\pm\gamma}$ are given by 
 
\begin{equation}
	\lambda^{\gamma}_\pm(\vert\mathbf{p}\vert, \beta)=\lambda^{-\gamma}_\pm(\vert\mathbf{p}\vert, \beta)=\frac{\rm{p}^2+\beta^2}{2}\pm \beta\vert \mathbf{p}\vert=E\pm\delta
\end{equation}
and the eigenvalues are real.

\vspace{05mm}

 The physical meaning of spin-orbit coupling is reflected in the expressions of eigenvalues where the split in the energy level for all $\beta\neq 0$ happens for a given value of momentum  $\mathbf{p}$. Such a splitting is also known as \textbf{Spin-band splitting} and this is a characteristic feature of \textbf{Bychkov-Rashba Effect}\cite{byrashba84}. The eigenvalues $E\pm\delta$ imply states with different helicities. The present discussion attempts to investigate the aspects of a  pseudo-Hermitian model Hamiltonian and bi-orthogonal eigen-spinors relating to the same effect .

The spinor eigenfunctions of $R^+_{\gamma}$ corresponding to $\lambda^{\gamma}_\pm(\vert\mathbf{p}\vert, \beta)$ are given by
$\vert\psi^{\gamma}_+(\mathbf{x}, \varphi_+) \rangle\rangle= 
\frac{1}{2\sqrt{2}\pi}\left(\begin{array}{c}
	e^{i\varphi_+} \\
	1
\end{array} \right)\otimes e^{i\mathbf{p}\cdot \mathbf{x}}$ and $\vert\psi_-^{\gamma}(\mathbf{x}, \varphi_-) \rangle\rangle= 
\frac{1}{2\sqrt{2}\pi}\left(\begin{array}{c}
-e^{i\varphi_-} \\
1
\end{array} \right)\otimes e^{i\mathbf{p}\cdot \mathbf{x}}$	
and those of $R^+_{-\gamma}$ corresponding to $\lambda^{-\gamma}_\pm$ are given by
$\vert\psi^{-\gamma}_+(\mathbf{x}, \varphi_-)\rangle\rangle = 
	\frac{1}{2\sqrt{2}\pi}\left(\begin{array}{c}
		1 \\
		e^{-i\varphi_-}
	\end{array} \right)\otimes e^{i\mathbf{p}\cdot \mathbf{x}}$ and $\vert\psi_-^{-\gamma}(\mathbf{x}, \varphi_+)\rangle\rangle = 
	\frac{1}{2\sqrt{2}\pi}\left(\begin{array}{c}
		-1 \\
		e^{-i\varphi_+}
	\end{array} \right)\otimes e^{i\mathbf{p}\cdot \mathbf{x}}$	
where,
$\varphi_{\pm}=\tan^{-1}\frac{\omega^2p_1\vert\mathbf{p}\vert\mp \gamma p_2^2}{\omega p_2\vert\mathbf{p}\vert\pm\gamma\omega p_1p_2}$. It is trivial to check that $\lambda^{-\gamma}_\pm=\lambda^{\gamma}_\pm$. This means ${\rm{Spec}}R_{\gamma}^+={\rm{Spec}}R_{-\gamma}^+$.

Let us define an inner-product $\mathcal{H}^\otimes\times\mathcal{H}^{\otimes}\rightarrow\mathbb{C}$ involving two spinor-states $\vert\psi\rangle\rangle=\vert A\rangle\otimes \vert \mathbf{p})$ and $\vert\phi\rangle\rangle=\vert B\rangle\otimes \vert \mathbf{p}^{\prime})$ in momentum representation
\begin{equation}
	\langle\langle\phi\vert \psi\rangle\rangle=\langle B\vert A\rangle\otimes(\mathbf{p}^{\prime}\vert\mathbf{p})=(b_1^{\star}a_1+b_2^{\star}a_2)\frac{1}{4\pi^2}\int e^{-i(\mathbf{p}^{\prime}-\mathbf{p})\cdot\mathbf{x}}d\mathbf{x}. 
\end{equation}

It is easy to check that $\langle\langle\psi^{\gamma}_+\vert \psi_-^{-\gamma}\rangle\rangle=0=\langle\langle\psi_-^{\gamma}\vert \psi_+^{-\gamma}\rangle\rangle$ in view of their column vectors of the respective states and hence the bi-orthogonality. \textbf{It is to be noted that the present bi-orthogonality involves two parameters $\varphi_\pm$ unlike Theorem-1 where a single parameter $\theta$ is involved}. 

\begin{remark}
	For $p_1=\pm p_2$, $\varphi_\pm$ is independent of momentum. \textbf{Furthermore the advantage of choosing $\vert\gamma\vert< 1$ to get a set of bi-orthogonal pair to construct the generators of $Cl_3(\mathbb{R})$ is reflected also in the expressions of bi-orthogonal states for $R_{\pm\gamma}$}. The expressions of $\varphi_\pm$ represent \textbf{real angles} for such a choice. The parameters $\varphi_\pm$ are related by the following equations

\begin{eqnarray}\label{flip}
 {\rm{(a)}}\:\varphi_\mp({\mathbf{p}}, \gamma)=\varphi_\pm({-\mathbf{p}}, \gamma)=\varphi_\pm({\mathbf{p}}, -\gamma)\nonumber\\
{\rm{(b)}}\: \varphi_\pm({-\mathbf{p}}, -\gamma)=\varphi_\pm({\mathbf{p}}, \gamma)\nonumber\\
 {\rm{(c)}}\:\varphi_\pm(-p_1, p_2, \gamma)+\varphi_\mp(p_1, p_2, \gamma)=0=\varphi_\pm(p_1, -p_2, \gamma)+\varphi_\pm(p_1, p_2, \gamma).
\end{eqnarray}
Equation-\ref{flip}-${\rm{(a)}}$ implies that finite parts of two spinors which are not bi-orthogonal partners interchange their parameters when the vector momentum $\mathbf{p}$ (or $\gamma$) is reversed keeping $\gamma$ (or vector momentum $\mathbf{p}$) fixed. Equation-\ref{flip}-${\rm{(b)}}$ supports the fact that the spinors retain their statuses under the simultaneous reversal of both $\mathbf{p}$ and $\gamma$. Equation-\ref{flip}-${\rm{(c)}}$ reveals that the reflection of momentum with respect to $X_2$-axis ($(p_1, p_2)\rightarrow (-p_1, p_2)$) and that with respect to $X_1$-axis    ($(p_1, p_2)\rightarrow (p_1, -p_2)$) do not produce identical effect unlike the Hermitian counterpart of the model.
\end{remark}

\begin{remark}\label{spinorvec}
The eigen-spinors obtained from the present Hamiltonian are of the form $\vert\psi\rangle\rangle=\left(\begin{array}{c}
		\psi_1 \\
		\psi_2
	\end{array} \right)=\frac{1}{2\sqrt{2}\pi}\left(\begin{array}{c}
		\pm 1 \\
		e^{\pm i\varphi}
	\end{array} \right) \otimes e^{i\mathbf{p}\cdot \mathbf{x}}$.
One can associate a vector $\mathbf{v}\in\mathbf{R}^3$  whose components can be identified with a spinor through the equations $\{v_j=\langle\langle\psi\vert\sigma_j\otimes \mathbf{1}\vert\psi\rangle\rangle : j=1, 2, 3\}$. In the present situation $v_3=0$ and hence $\mathbf{v}$ is a vector in $X_1X_2$-plane. This vector depends on $\mathbf{p}$ and $\gamma$ in a very intricate way unlike the Hermitian case where, it depends only on the ratio of $p_1$ and $p_2$. This means for a fixed ratio of $p_1$ and $p_2$ the alignment of this vector can be varied by changing the value of $\gamma$.  This fact seems to have some theoretical and experimental implications in the study of many electron systems showing spin precession, spin damping, spin polarization and relaxation mechanisms.
\end{remark}

\begin{remark}
It is to be mentioned that for any time-evolving state the current density operator as derived from respective time dependent Schr\"odinger equations for $R_{\pm\gamma}^+$ gives us a $\gamma$-deformed expression given by $\mathbf{J}=\frac{1}{2i}\mathbf{1}_2\otimes[\mathbf{P}\vert\mathbf{x})(\mathbf{x}\vert+\vert\mathbf{x})(\mathbf{x}\vert\mathbf{P}]+\beta [\omega^{-1}e_1\otimes\vert\mathbf{x})(\mathbf{x}\vert \mathbf{u}_2-e_2\otimes\vert\mathbf{x})(\mathbf{x}\vert \mathbf{u}_1]$, where, $\mathbf{P}=P_1\mathbf{u}_1+P_2\mathbf{u}_2$ being the Cartesian vector momentum operator and $\mathbf{u}_{1, 2}$ are unit vectors along the respective co-ordinate directions. The current density $\langle\mathbf{J}\rangle=\langle\langle\psi\vert\mathbf{J}\psi\rangle\rangle$ follows the continuity relation $\mathbf{\nabla}\cdot\langle\mathbf{J}\rangle+\partial_t\rho=0$ with $\rho=\psi^\star\psi$.
\end{remark}

Let us explore the role of the interaction term concerning bi-orthogonality. The following proposition is immediate. 

\begin{prop}
Given two isospectral Hamiltonians $H$ and $H^{\dagger}$ $(H\neq H^{\dagger})$ with eigenvalue equations

\begin{equation}\label{isospec}
 {\rm{(a)}}. H\vert\xi_\pm\rangle=(E\pm\delta)\vert\xi_\pm\rangle\:\:{\rm{and}}\:\: {\rm{(b)}}. H^\dagger\vert \zeta_\pm\rangle=(E\pm\delta)\vert\zeta_\pm\rangle,
 \end{equation}
 
 $E, \delta\in\mathbb{R}$,  $\langle\zeta_-\vert\xi_+\rangle=0=\langle\zeta_+\vert\xi_-\rangle$ for $\delta\neq 0$
\end{prop}
\textbf{Proof} : Left multiplying equation-\ref{isospec}-(a) by $\langle\zeta_\mp\vert$ and right multiplying the adjoint of equation-\ref{isospec}-(b) by $\vert\xi_\mp\rangle$ we get
\begin{equation}
 {\rm{(a)}}. \langle\zeta_\mp\vert H\vert\xi_\pm\rangle=(E\pm\delta)\langle\zeta_\mp\vert \xi_\pm\rangle\:\:{\rm{and}}\:\:{\rm{(b)}}. \langle\zeta_\pm\vert H\vert\xi_\mp\rangle=(E\pm\delta)\langle\zeta_\pm\vert \xi_\mp\rangle.
\end{equation}
Subtracting the second from the first  we get $2\delta\langle\zeta_-\vert \xi_+\rangle=0=-2\delta\langle\zeta_+\vert \xi_-\rangle$. Noting $\delta\neq 0$, $\langle\zeta_-\vert \xi_+\rangle=0=\langle\zeta_+\vert \xi_-\rangle$. $\square$

\vspace{0.5cm}

\begin{remark}
The above proposition reveals the fact that the pair of Hamiltonians so chosen can produce bi-orthogonal set and this is exactly the situation with $R^+_{\pm\gamma}$. In addition to that the very property of bi-orthogonality is validated due to non-zero contribution of the non-Hermitian spin-orbit term.
\end{remark}

One can also construct the bi-orthogonal spinor projection operators
\begin{eqnarray}
	\Pi_1^+=\frac{\vert\psi_+^{\gamma}\rangle\rangle\langle\langle \psi_+^{-\gamma}\vert}{\langle\langle \psi_+^{-\gamma}\vert\psi_+^{\gamma}\rangle\rangle}\nonumber\\
	=\frac{1}{e^{i\varphi_+}+e^{i\varphi_-}}\left(\begin{array}{cc}
	e^{i\varphi_+} &	e^{i(\varphi_++\varphi_-)} \\
	1 & e^{i\varphi_-}	
	\end{array} \right)\otimes \mathbf{1}_\infty
\end{eqnarray}
and
\begin{eqnarray}
	\Pi_2^+=\frac{\vert\psi_-^{\gamma}\rangle\rangle\langle\langle \psi_-^{-\gamma}\vert}{\langle\langle \psi_-^{-\gamma}\vert\psi_-^{\gamma}\rangle\rangle}\nonumber\\
	=\frac{1}{e^{i\varphi_+}+e^{i\varphi_-}}\left(\begin{array}{cc}
		e^{i\varphi_-} &	-e^{i(\varphi_++\varphi_-)} \\
		-1 & e^{i\varphi_+}	
	\end{array} \right)\otimes \mathbf{1}_\infty,
\end{eqnarray}
where, $\int\vert\mathbf{p})( \mathbf{p}\vert d \mathbf{p}=\mathbf{1}_\infty$ implies formal unit operator.
 It can be verified that $\Pi_1^++\Pi_2^+=\mathbf{1}_2\otimes \mathbf{1}_\infty$ and $\Pi_1^+\Pi_2^+=0$. Choosing a different ansatz like $\vert\phi(\mathbf{x})\rangle\rangle = 
 \frac{1}{2\pi}\left(\begin{array}{c}
 	b_1 \\
 	b_2
 \end{array} \right)\otimes e^{-i\mathbf{p}\cdot \mathbf{x}}$ we get similar expressions of eigenvalues for $R^+_{\gamma}$ and $R^+_{-\gamma}$ respectively i. e.;
\begin{equation}
	\mu^{\gamma}_\pm(\vert\mathbf{p}\vert, \beta)=\mu^{-\gamma}_\pm(\vert\mathbf{p}\vert, \beta)=\frac{\rm{p}^2+\beta^2}{2}\pm \beta\vert\mathbf{p}\vert.
\end{equation}	

The eigen-spinors of $R^+_{\gamma}$ corresponding to $\mu^{\gamma}_{\pm}$ are given by $\vert\phi_{\pm}^{\gamma}(\mathbf{x})\rangle\rangle = 
\frac{1}{2\sqrt{2}\pi}\left(\begin{array}{c}
	e^{i\varphi_\mp} \\
	1
\end{array} \right)\otimes e^{-i\mathbf{p}\cdot \mathbf{x}}$ and those of $R^+_{-\gamma}$ corresponding to $\mu_\pm^{-\gamma}$ are $\vert\phi_{\pm}^{-\gamma} (\mathbf{x})\rangle\rangle= 
\frac{1}{2\sqrt{2}\pi}\left(\begin{array}{c}
-1 \\
e^{-i\varphi_\pm}
\end{array} \right)\otimes e^{-i\mathbf{p}\cdot \mathbf{x}}$. One can verify that $\langle\langle\phi^{\gamma}_+\vert \phi_-^{-\gamma}\rangle\rangle=0=\langle\langle\phi_-^{\gamma}\vert \phi_+^{-\gamma}\rangle\rangle$.

\begin{remark}
In order to avoid the problem with transition probabilities within the states of the Hamiltonian (for example $\langle\langle\psi_+^\gamma\vert\psi_-^\gamma\neq 0$) it is suggestive, following \cite{brody14},  to consider the so called associated state corresponding to a given state $\vert\psi\rangle\rangle=c_+\vert\psi_+^\gamma\rangle\rangle+c_-\vert\psi_-^\gamma\rangle\rangle$ given by 
\begin{equation}
\vert\breve{\psi}\rangle\rangle=c_+\vert\psi_+^{-\gamma}\rangle\rangle+c_-\vert\psi_-^{-\gamma}\rangle\rangle
\end{equation} 
and the expectation value of any operator $K$ is given by $\langle\langle K\rangle\rangle=\frac{\langle\langle\breve{\psi}\vert K\vert\psi\rangle\rangle}{\langle\langle\breve{\psi}\vert\psi\rangle\rangle}$
\end{remark}

\vspace{0.5cm}

\textbf{Comments on Completeness} : 
The type of functions for example, $f_{\mathbf{p}}(\mathbf{x})=(\mathbf{x}\vert\mathbf{p)}=\frac{1}{2\pi}e^{i\mathbf{p}\cdot \mathbf{x}}$ does not belong to a Hilbert space of square integrable functions $L^2(\mathbb{R}^2)$ though the set $\{f_{\mathbf{p}}(\mathbf{x})\}$ can be viewed as complete set of eigenfunctions of  the operator $\mathbf{P}$ ( as the generator of  translation) or more generally $U_{\mathbf{h}}=e^{i\mathbf{P}\cdot\mathbf{h}}$ in global sense defined by the action $U_{\mathbf{h}}g(\mathbf{x})=g(\mathbf{x}+\mathbf{h})$. Choosing $g(\mathbf{x})=f_{\mathbf{p}}(\mathbf{x})$ we get $U_{\mathbf{h}}f_{\mathbf{p}}(\mathbf{x})=e^{i\mathbf{p}\cdot\mathbf{h}}f_{\mathbf{p}}(\mathbf{x})$. This means $f_{\mathbf{p}}(\mathbf{x})$ is an eigenfunction of $U_{\mathbf{h}}$ with the eigenvalue $e^{i\mathbf{p}\cdot\mathbf{h}}$. To show that $\{f_{\mathbf{p}}(\mathbf{x}) : \mathbf{p}\in\mathbf{R}^2\}$ is complete, let us consider the following Fourier Transform and Inverse Fourier Transform

\begin{eqnarray}\label{fourier}
{\rm{(a)}}\:\:g(\mathbf{x})=(\mathbf{x}\vert g)=\int (\mathbf{x}\vert\mathbf{p})(\mathbf{p}\vert g)d\mathbf{p}=\frac{1}{2\pi}\int \bar{g}(\mathbf{p})e^{i\mathbf{p}\cdot\mathbf{x}}d\mathbf{p}\:\:\:{\rm{and}}\nonumber\\
{\rm{(b)}}\:\:\bar{g}(\mathbf{p})=\frac{1}{2\pi}\int g(\mathbf{x})e^{-i\mathbf{p}\cdot\mathbf{x}}d\mathbf{x}
\end{eqnarray}

Applying $U_{\mathbf{h}}$ on both sides of  equation \ref{fourier} a
\begin{eqnarray}\label{spectral}
U_{\mathbf{h}}g(\mathbf{x})=\frac{1}{2\pi}\int e^{i\mathbf{p\cdot\mathbf{h}}}\bar{g}(\mathbf{p})e^{i\mathbf{p\cdot\mathbf{x}}}d\mathbf{p}\nonumber\\
=\frac{1}{2\pi}\int e^{i\mathbf{p\cdot\mathbf{h}}}\left[\frac{1}{2\pi}\int g(\mathbf{x}^{\prime})e^{-i\mathbf{p}\cdot\mathbf{x}^{\prime}}d\mathbf{x}^{\prime}\right]e^{i\mathbf{p\cdot\mathbf{x}}}d\mathbf{p}
\end{eqnarray}

Equation-\ref{spectral} is the infinite dimensional interpretation of the theorem involving unitary operator that, for finite dimension, states that  if $\{\vert v_j\rangle : j=1,2,\cdots, n \}$ are the eigen-vectors of the unitary operator $U$ with eigenvalues $\{\lambda_j : j=1, 2,\cdots, n\}$ then for any vector $\vert w\rangle$,  $U\vert w\rangle=\sum_{j=1}^n\lambda_j\langle v_j\vert w\rangle\vert v_j\rangle$ and the set $\{\vert v_j\rangle\}$ is complete \cite{gelfand64}. In equation-\ref{spectral} the completeness is ensured by the Plancherel relation $\int\vert g(\mathbf{x})\vert^2d\mathbf{x}=\int\vert \bar{g}(\mathbf{p})\vert^2d\mathbf{p}$. {\footnote{The completeness argument can be grounded more rigorously following the notion of \textbf{Rigged Hilbert Space} where the vectors in a given space $\mathcal{H}$ are considered with a topological refinement $\mathcal{H}_{\prec}\subset\mathcal{H}\subset \mathcal{H}_\prec^\times$, $\mathcal{H}_\prec^\times$ being the anti-dual of $\mathcal{H}_\prec$. This triplet of spaces is known as \textbf{Gelfand Triple}. The so-called \textbf{Nuclear Spectral Theorem} due to Gelfand \cite{gelfand64} ensures the completeness in this context. }}

\section{Pseudo-Hermiticity, fermionic time reversal and Clifford algebra}\label{sectionpseudo}

\subsection{\textbf{Pseudo-Hemiticity, time reversal and and an analogue of Kramers' theorem}}

\begin{dfn}
The fermionic time reversal operator $\mathcal{T}$ is defined by  the following action on spinor state $\vert\psi\rangle\rangle$ :
\begin{eqnarray}
\mathcal{T}\vert\psi\rangle\rangle=\mathcal{T}\left(\begin{array}{c}
	\psi_1 \\
	\psi_2
\end{array} \right)=\mathcal{T}\left(\begin{array}{c}
	a_1 \\
	a_2
\end{array} \right)\otimes e^{i\mathbf{p}\cdot\mathbf{x}}=\left(\begin{array}{c}
	-a_2^{\star} \\
	a_1^{\star}
\end{array} \right)\otimes e^{-i\mathbf{p}\cdot\mathbf{x}}\nonumber\\
=\left(\begin{array}{c}
	-\psi_2^\star \\
	\psi_1^\star
\end{array} \right)\:\:\:{\rm{and}}\:\:\:
\mathcal{T}^2\vert\psi\rangle\rangle=-\vert\psi\rangle\rangle.
\end{eqnarray}. 
\end{dfn}

Choosing $\mathcal{T}=e_{13}\mathcal{T}_0$, where, $\mathcal{T}_0\left(\begin{array}{c}
	\psi_1 \\
	\psi_2
\end{array} \right)=\left(\begin{array}{c}
	\psi_1^\star \\
	\psi_2^\star
\end{array} \right)$ {\footnote{It may be noted that considering $\varphi\rightarrow\varphi+\pi$ as the consequence of time reversal the vectors corresponding to a bi-orthogonal pair (section-\ref{biortho}, remark-\ref{spinorvec}) are time reversed versions of each other. }} and using the facts $\mathcal{T}^{-1}\sigma_m^{\gamma}\mathcal{T}=-\sigma_m^{-\gamma}$, $\mathcal{T}^{-1}P_m\mathcal{T}=-P_m$ for all $m=1, 2, 3$ and $\mathcal{T}^{-1} i \mathcal{T}=-i$, it can be readily verified that under the action of $\mathcal{T}$, $R^+_{\pm\gamma}$ is $\mathcal{T}$-pseudo-Hermitian that is
\begin{equation}
	\mathcal{T}^{-1}R^+_{\pm\gamma}\mathcal{T}=(R^+_{\pm\gamma})^{\dagger}\:\:\:{\rm{and}}\:\:{\rm{furthermore}}\:\:(R^+_{\pm\gamma})^{\dagger}=R^+_{\mp\gamma}.
\end{equation}  

\begin{theorem}\label{schrodinger}
Given the time-dependent Schr\"odinger equation
\begin{eqnarray}
H\vert\psi(t)\rangle=i\partial_t\vert\psi(t)\rangle
\end{eqnarray}
satisfied by $\psi(t)\rangle$, $\mathcal{T}\vert\psi\rangle$ satisfies the time reversed Schr\"odinger equation with the Hamiltonian $H^\dagger=\mathcal{T}^{-1}H\mathcal{T}$.
\end{theorem}

\textbf{Proof}
Effecting time reversal $\mathcal{T}$ on both sides one can write
\begin{eqnarray}
\mathcal{T}H\mathcal{T}^{-1}\mathcal{T}\vert\psi(t)\rangle &=&\mathcal{T}i\mathcal{T}^{-1}\mathcal{T}\partial_t\mathcal{T}^{-1}\mathcal{T}\vert\psi(t)\rangle\nonumber\\
{\rm{or}}\:\:\mathcal{T}^{-1}H\mathcal{T}\vert\psi(t)\rangle &=&-i\partial_{-t}\mathcal{T}\vert\psi(t)\rangle\:\:{\rm{using}}\:\:\mathcal{T}^{-1}=-\mathcal{T}\nonumber\\
{\rm{or}}\:\:H^\dagger\vert\mathcal{T}\psi\rangle &=&i\partial_t\vert\mathcal{T}\psi\rangle\:\:\square
\end{eqnarray}.

\vspace{0.5cm}

The following properties of $\mathcal{T}$ are very crucial

\begin{theorem}
$\mathcal{T}$ is anti-unitary, norm preserving and $\vert\psi\rangle$ and $\mathcal{T}\vert\psi\rangle$ are orthogonal.
\end{theorem}

\textbf{Proof}
Considering the obvious result
\begin{eqnarray}\label{timerev}
\langle\mathcal{T}\psi\vert\mathcal{T}\phi\rangle=\psi_1\phi_1^\star+\psi_2\phi_2^\star=\langle\phi\vert\psi\rangle.
\end{eqnarray}
 Putting $\vert\phi\rangle=\vert\psi\rangle$ in the equation-\ref{timerev}, $\langle\mathcal{T}\psi\vert\mathcal{T}\psi\rangle=\langle\psi\vert\psi\rangle$. This proves $\mathcal{T}$ is norm-preserving.

On the other hand, putting $\vert \phi\rangle=\mathcal{T}\vert\psi\rangle$ in equation-\ref{timerev} the l. h. s. gives $\langle\mathcal{T}\psi\vert\mathcal{T}^2\psi\rangle=-\langle\mathcal{T}\psi\vert\psi\rangle$ while the r. h. s. gives $\langle\phi\vert\psi\rangle=\langle\mathcal{T}\psi\vert\psi\rangle$ or $\langle\mathcal{T}\psi\vert\psi\rangle=0\:\:\square$.

 \vspace{.5cm}

\textbf{Comments on an analogue of Kramers' theorem}: It is to be mentioned that neither the operators $\{e_1^{\gamma}, e_2^{\gamma}\}$ nor the components of momentum operator $\{P_1, P_2\}$ possess  the said $\mathcal{T}$-pseudo-Hermiticity property at their individual level. However, the spin-orbit interaction term adopts this property.\textbf{ In Hermitian quantum mechanics Kramers' theorem tells that for fermionic systems with half-integer total spin where time reversal symmetry (TRS) is present, all energy levels are doubly degenerate}. Since, the time reversal symmetry implies the commutativity of $\mathcal{T}$ and the Hamiltonian and in the present case the operator $\mathcal{T}$ does not commute with the Hamiltonians $R^\pm_{\gamma}$, such a system does not deserve the conventional sense of Kramers' degeneracy theorem. However, an alternative version of the same can be possible in the present context through the following observation.

Noting  $R^+_{\pm\gamma}$ to be iso-spectral  and eigenvalues are independent of $\gamma$, we can make the following inferences.

\begin{inference}\label{inf}
1. If $\vert\psi_{\pm}^\gamma\rangle\rangle$ is an eigenvector of $R^+_\gamma$, $\mathcal{T}\vert\psi_\pm^{\gamma}\rangle\rangle$ is an eigenvector of $(R^+_\gamma)^\dagger$ and they are mutually orthogonal or in other words, \textbf{the finite part of the time reversed state vector is proportional to that of  the bi-orthogonal partner of the state vector itself.} 

2. It is to be noted that  $\mathcal{T}\vert\psi_\pm^\gamma\rangle=(-1)^n\vert\psi_\mp^{-\gamma}\rangle$ with $n=0\:\:\:{\rm{or}}\:\:\:1$ depending on the choice of eigenvectors. Now, $\vert\psi_{\pm}^\gamma\rangle\rangle$ and $\mathcal{T}\vert\psi_{\mp}^\gamma\rangle\rangle$ are eigenvectors of $R^+_\gamma$ and $(R^+_\gamma)^\dagger$ corresponding to  eigenvalues $E\pm\delta$ and $E\mp\delta$. 
\end{inference}

Similar observation holds for $\vert\phi^{\pm\gamma}_\pm\rangle\rangle$ and $\mathcal{T}\vert\phi^{\pm\gamma}_\pm\rangle\rangle$ as well. Inference-\ref{inf} may be considered as an analogue of Kramers' theorem in bi-orthogonal quantum mechanics however still conjectural in sense.

\subsection{\textbf{Time reversal and Clifford algebra}}

\subsubsection{\textbf{Involutions and spinors as left ideal of $Cl_3$}}

\begin{dfn}
	A unary operation $\sharp$ acting on an element $\mathbf{U}$ is said to be an involution if $(\mathbf{U}^{\sharp})^{\sharp}=\mathbf{U}$. An involution may be an automorphism or an anti-automorphism.
\end{dfn}

For an arbitrary element $\mathbf{U}\in Cl_3$ consisting of a scalar $\langle \mathbf{U}\rangle_0$, a vector $\langle \mathbf{U}\rangle_1$, a bi-vector $\langle \mathbf{U}\rangle_2$ and a tri-vector $\langle \mathbf{U}\rangle_3$  following three types of involution are possible.

\begin{eqnarray*}
	\widehat{\mathbf{U}}=\langle \mathbf{U}\rangle_0-\langle \mathbf{U}\rangle_1+\langle \mathbf{U}\rangle_2-\langle \mathbf{U}\rangle_3:\:\:{\rm{grade}}\:\:{\rm{inversion}}\:\:{\rm{(automorphism)}}\nonumber\\ 
	\widetilde{\mathbf{U}}=\langle \mathbf{U}\rangle_0+\langle \mathbf{U}\rangle_1-\langle \mathbf{U}\rangle_2-\langle \mathbf{U}\rangle_3 : \:\:{\rm{reversion}}\:\:{\rm{(antiautomorphism)}}\nonumber\\
	\overline{\mathbf{U}}=\langle \mathbf{U}\rangle_0-\langle \mathbf{U}\rangle_1-\langle \mathbf{U}\rangle_2+\langle \mathbf{U}\rangle_3 : \:\:{\rm{Clifford}}\:\:{\rm{conjugation}}\:\:{\rm{(antiautomorphism)}}.
\end{eqnarray*} 

 We shall make use of these involutions to observe various inner-products on $\mathcal{S}$, a minimal ideal of $Cl_3$. Considering $\mathbf{U}=\left(\begin{array}{cc}
	u_{11} & u_{12}\\
	u_{21} & u_{22}
\end{array} \right)\in Cl_3$ in terms of conventional $Cl_3$ bases $\mathcal{G}=\{1; e_1, e_2, e_3; e_{12}, e_{23}, e_{31}; e_{123}\}$ the above involutions are equivalent to the following expressions

\begin{eqnarray}
	\widehat{\mathbf{U}}=\left(\begin{array}{cc}
		u_{22}^{\star} & -u_{21}^{\star}\\
		-u_{12}^{\star} & u_{11}^{\star}
	\end{array} \right),\:\:\widetilde{\mathbf{U}}=\left(\begin{array}{cc}
		u_{11}^{\star} & u_{21}^{\star}\\
		u_{12}^{\star} & u_{22}^{\star}
	\end{array} \right)	\:\:{\rm{and}}\nonumber\\
	\overline{\mathbf{U}}=\left(\begin{array}{cc}
		u_{22} & -u_{12}\\
		-u_{21} & u_{11}
	\end{array} \right).	
\end{eqnarray}

Let us write the following matrices with the help of $\mathcal{G}^{\gamma}$ and $\breve{\mathcal{G}}^{\gamma}=\{{1}, \breve{e}^{\gamma}_1, \breve{e}^{\gamma}_2, \breve{e}^{\gamma}_3, \breve{e}^{\gamma}_{12}, \breve{e}^{\gamma}_{23}, \breve{e}^{\gamma}_{31}, \breve{e}^{\gamma}_{123} \}=\{\mathbf{1}_2, -\tilde{\sigma}_1^{\gamma}, \tilde{\sigma}_2^{\gamma}, -\tilde{\sigma}_3^{\gamma}, i\tilde{\sigma}_3^{\gamma}, i\tilde{\sigma}_1^{\gamma}, i\tilde{\sigma}_2^{\gamma}, -i\mathbf{1}_2\}$ (where, the later may be called the set of \textbf{time reversed generators} defined by $\breve{g}= \mathcal{T}^{-1}g\mathcal{T}$ for any generator $g\in Cl_3(\mathbb{R})$).

\begin{eqnarray}
	g_0=\frac{1}{2}\mathbf{1}_2+\frac{\omega}{4}(e_3^{\gamma}-\breve{e}_3^{\gamma})=\left(\begin{array}{cc}
		1 & 0\\
		0 & 0
	\end{array} \right)\:\:
g_1=\frac{1}{2}e_2^{\gamma}+\frac{\omega}{4}(e_{23}^{\gamma}+\breve{e}_{23}^{\gamma})=\left(\begin{array}{cc}
	0 & 0\\
	i & 0
\end{array} \right)\nonumber\\
g_2=\frac{1}{2}(e_{31}^{\gamma})-\frac{\omega}{4}(e_1^{\gamma}-\breve{e}_1^{\gamma})=\left(\begin{array}{cc}
	0 & 0\\
	-1 & 0
\end{array} \right)\nonumber\\
g_3=\frac{1}{2}e_{123}+\frac{\omega}{4}(e_{12}^{\gamma}+\breve{e}_{12}^{\gamma})=\left(\begin{array}{cc}
	i & 0\\
	0 & 0
\end{array} \right).
\end{eqnarray}

With $\{\zeta_0, \zeta_1, \zeta_2, \zeta_3\}$ being real functions, we can write an element $\Psi=\sum_{j=0}^{3}\zeta_jg_j$ of the left ideal  $\mathcal{S}$ (of $Cl_3$) in terms of the generators $\mathcal{G}_s=\{g_j\vert j=0, 1, 2, 3\}$ derived from bi-orthogonal system. Identifying $\psi_1=\zeta_0+i\zeta_3$ and $\psi_2=-\zeta_2+i\zeta_1$, with $\zeta_0=\vert a\vert\cos(\mathbf{p}\cdot\mathbf{x}+\theta_a) $,  $\zeta_1=\vert b\vert\sin(\mathbf{p}\cdot\mathbf{x}+\theta_b) $, $\zeta_2=-\vert b\vert\cos(\mathbf{p}\cdot\mathbf{x}+\theta_b) $ and $\zeta_3=\vert a\vert\sin(\mathbf{p}\cdot\mathbf{x}+\theta_a) $ $(\tan\theta_a=a_2/a_1, \tan\theta_b=b_2/b_1)$ a typical spinor state $\psi=\left(\begin{array}{c}
	\psi_1 \\
	\psi_2
\end{array} \right)\in\mathbf{C}^2\otimes\mathcal{H}_0$ can be cast into $2\times 2$ matrix form ${\Psi_{\rm{Cliff}}}=\left(\begin{array}{cc}
\psi_1 & 0\\
\psi_2 & 0
\end{array} \right)\in M(2, \mathbb{C})g_0$, where, $M(2, \mathbb{C})$ is the space of $2\times 2$ matrices over $\mathbb{C}$, $g_0=\left(\begin{array}{cc}
1 & 0\\
0 & 0
\end{array} \right)$ is an idempotent $g_0^2=g_0$ and $M(2, \mathbb{C})g_0\cong \mathbf{C^2}$. $\psi\in\mathcal{S}=M(2, \mathbb{C})g_0\subset {Cl}_3$ is the minimal left ideal of ${Cl}_3$ in the sense that any $u\in {Cl}_3$ and $\psi\in\mathcal{S}$, $u\psi\in \mathcal{S}$. 

The effect of time reversal in $\mathcal{S}$ itself can also be understood in terms of a unary operation $'\flat'$ which is equivalent to \textbf{basis flip} as given below
\begin{equation}
\left.\begin{array}{c}
\mathbf{1}_2\rightarrow e_{13}\\
e_{13}\rightarrow -\mathbf{1}_2
\end{array} \right\}, \left.\begin{array}{c}
e_1\rightarrow -e_{3}\\
e_{3}\rightarrow e_1
\end{array} \right\}, \left.\begin{array}{c}
e_2\rightarrow e_{123}\\
e_{123}\rightarrow -e_2
\end{array} \right\}\:\:{\rm{and}}\:\:\left.\begin{array}{c}
e_{12}\rightarrow -e_{23}\\
e_{23}\rightarrow e_{12}
\end{array} \right\}.
\end{equation}

 This makes $\mathbf{U}^\flat=\left(\begin{array}{cc}
-u^\star_{21} & -u^\star_{22}\\
u^\star_{11} & u^\star_{12}
\end{array} \right)$ and $(\mathbf{U}^\flat)^\flat=-\mathbf{U}$ resulting to $\Psi_{\rm{Cliff}}^{\flat}$, the time reversed version of $\Psi_{\rm{Cliff}}$ and also an element of the left ideal $\mathcal{S}$. It can be verified that $(\Psi_{\rm{Cliff}}^{\flat})^{\flat}=-\Psi_{\rm{Cliff}}$ and this fact is consistent with the anti-involution property of fermionic time reversal.

\begin{remark}
It is to be noted that the generalized Clifford momenta $\{\wp^A_\gamma, \wp^B_\gamma\}$ defined in section-\ref{cliffmomentum} do not have reflection symmetry under time-reversal like ordinary Cartesian momenta $\{-i\partial_1, -i\partial_2, -i\partial_3\}$. It is obvious to see that $\mathcal{T}^{-1}\wp_\gamma^{A, B}\mathcal{T}=\wp_{-\gamma}^{A, B}$, the later being represented in terms of time-reversed bases. This means $\wp_\gamma^{A, B}$ are neither Hermitian nor $\mathcal{T}$-pseudo-Hermitian in general.
\end{remark}

\subsubsection{\textbf{Involutions and inner-products}}

The aim of this section is to interpret the inner-product as a consequence of various Clifford involutions and their relations to bi-orthogonality and time-reversal.

\begin{prop} 
Let's  consider a division ring $\mathcal{R}=g_0Cl_3g_0=\left\{\left(\begin{array}{cc}
	c & 0  \\
	0 & 0
\end{array} \right) : c\in\mathbb{C}\right\}$  with $g_0\in Cl_3$ being the primitive and $\Psi_{\rm{Cliff}}=\mathbf{A}e^{\pm i\mathbf{p}\cdot\mathbf{x}},\Phi_{\rm{Cliff}}=\mathbf{B}e^{\pm i\mathbf{p}\cdot\mathbf{x}}\in\mathcal{S}$ with $\mathbf{A}=\left(\begin{array}{cc}
 		a_{1} & 0\\
 		a_{2} & 0
 	\end{array} \right)$ and $\mathbf{B}=\left(\begin{array}{cc}
 		b_{1} & 0\\
 		b_{2} & 0
 	\end{array} \right)$

I. The map $\mathcal{C}_1 : = \widetilde{\Psi}_{\rm{Cliff}}\Phi_{\rm{Cliff}} $ defines a map $\mathcal{S}\times \mathcal{S}\rightarrow \mathcal{R}$ and $\langle\langle{\Psi}_{\rm{Cliff}},\Phi_{\rm{Cliff}}\rangle\rangle_1=\tr\tilde{\mathbf{A}}{\mathbf{B}}\frac{1}{4\pi^2}\int e^{i(\mathbf{p}-\mathbf{p}^{\prime})\cdot\mathbf{x}}d\mathbf{x}$ is equal to the inner-product $\langle\langle\psi\vert\phi\rangle\rangle$ ( for which $\langle\langle{\Psi}_+^{\gamma}, \Psi_-^{-\gamma}\rangle\rangle_1=0={\langle\langle{{\Psi}_+^{\gamma}}, {\Psi}_-^{-\gamma}\rangle\rangle_1}$ and $\langle\langle{\Psi}_-^{\gamma}, \Psi_+^{-\gamma}\rangle\rangle_1=0={\langle\langle{{\Psi}_-^{\gamma}},{\Psi}_+^{-\gamma}\rangle\rangle}_1$). 

II. The anti-unitarity of time reversal is equivalent to the claim $\langle\langle{\Phi}^{\flat}_{\rm{Cliff}},\Psi^{\flat}_{\rm{Cliff}}\rangle\rangle_1=\langle\langle{\Psi}_{\rm{Cliff}},\Phi_{\rm{Cliff}}\rangle\rangle_1$. 

III. The map  $\mathcal{C}_2 : \mathcal{S}\times \mathcal{S}\rightarrow \mathcal{R} : =e_{31}\overline{\Psi}_{\rm{Cliff}}\Phi^{\flat}_{\rm{Cliff}}$ defines an  inner-product $\langle\langle{\Psi}_{\rm{Cliff}},\Phi_{\rm{Cliff}}\rangle\rangle_2=\tr e_{31}\overline{\mathbf{A}}{\mathbf{B}^{\flat}}\frac{1}{4\pi^2}\int e^{\pm i(\mathbf{p}-\mathbf{p}^{\prime})\cdot\mathbf{x}}d\mathbf{x}$ which justifies the bi-orthogonality $\langle\langle{\Psi}_+^{\gamma}, \Psi_-^{-\gamma}\rangle\rangle_2=0={\langle\langle{{\Psi}_-^{\gamma}},{\Psi}_+^{-\gamma}\rangle\rangle}_2$.
\end{prop}

\textbf{Proof} : I.  The first part is obvious. For the second part we observe

 \begin{eqnarray}
 	\langle\langle{\Psi}_{\rm{Cliff}}, \Phi_{\rm{Cliff}}\rangle\rangle_1 &=&\tr\tilde{\mathbf{A}}{\mathbf{B}}\frac{1}{4\pi^2}\int e^{\pm(\mathbf{p}-\mathbf{p}^{\prime})\cdot\mathbf{x}}d\mathbf{x}\nonumber\\
 	&=&\tr\left(\begin{array}{cc}
 		a_{1}^{\star} & a_{2}^{\star}\\
 		0 & 0
 	\end{array} \right)	\left(\begin{array}{cc}
 	b_{1} & 0\\
 	b_{2} & 0
 \end{array} \right)\frac{1}{4\pi^2}\int e^{\pm(\mathbf{p}-\mathbf{p}^{\prime})\cdot\mathbf{x}}d\mathbf{x}\nonumber\\
 &=& (a_{1}^{\star}b_{1}+a_{2}^{\star}b_{2} )\frac{1}{4\pi^2}\int  e^{\pm(\mathbf{p}-\mathbf{p}^{\prime})\cdot\mathbf{x}}d\mathbf{x}=\langle\langle\psi\vert\phi\rangle\rangle.
\end{eqnarray}

$\langle\langle{\Psi}_{\rm{Cliff}}, \Phi_{\rm{Cliff}}\rangle\rangle_1=0$ implies orthogonality (or bi-orthogonality in our case). 

II. The anti-unitarity of time reversal is obvious by definition.

III. Obvious by definition. $\square$
 
\vspace{0.5cm} 

\begin{remark}
The transformations that leave the product $\widetilde{\Psi}_{\rm{Cliff}}\Phi_{\rm{Cliff}}$ invariant constitute the group
$\mathfrak{G}=\{u : \tilde{u}u=\mathbf{1}_{2\times 2}\}$ which is isomorphic to the group of unitary matrices $\mathcal{U}(2)=\{u\in{\rm{Mat}}(2, \mathbb{C}) : u^{\dagger}u=\mathbf{1}_{2\times 2}\}$. Similarly, the invariance group for the product $e_{31}\overline{\Psi}_{\rm{Cliff}}\Phi^{\flat}_{\rm{Cliff}}$ is given by $\mathfrak{G}^{\prime}=\{u : \overline{u}u^{\flat}=\mathbf{1}_{2\times 2}\}$.
\end{remark}

\section{SUSY and pseudo-SUSY structure}\label{sectionsusy}
In view of the factorizability of the Hamiltonians $R^{\pm}_{\pm\gamma}$ let us introduce the so-called super-charge operators
\begin{equation}
\Theta^+_{\pm\gamma}=\frac{1}{\sqrt{2}}\left(\begin{array}{cc}
\mathbf{0}_{2\times 2} & \wp^B_{\pm\gamma} \\
\mathbf{0}_{2\times 2} & \mathbf{0}_{2\times 2}
\end{array} \right)\:\:{\rm{and}}\:\:\Theta^-_{\pm\gamma}=\frac{1}{\sqrt{2}}\left(\begin{array}{cc}
\mathbf{0}_{2\times 2} & \mathbf{0}_{2\times 2}\\
\wp^A_{\pm\gamma}  & \mathbf{0}_{2\times 2}
\end{array} \right)
\end{equation}
with $(\Theta^\pm_{\pm\gamma})^2=0.$
 These lead to two SUSY Hamiltonians given by
 \begin{equation}\label{susyhamil}
 H^{\rm{SUSY}}_{\pm\gamma}=\{\Theta^+_{\pm\gamma}, \Theta^-_{\pm\gamma}\}=\left(\begin{array}{cc}
R^+_{\pm\gamma} & \mathbf{0}_{2\times 2}\\
\mathbf{0}_{2\times 2}  & R^-_{\pm\gamma} 
\end{array} \right)
 \end{equation}
along with the fact $[H^{\rm{SUSY}}_{\pm\gamma}, \Theta^+_{\pm\gamma}]=0=[H^s_{\pm\gamma}, \Theta^-_{\pm\gamma}]$. It is to be noted that the Hilbert space $\mathcal{H}^{\rm{SUSY}}$ on which $H^{\rm{SUSY}}_{\pm\gamma}$ acts admits the so called $\mathbb{Z}_2$ grading i. e.; $\mathcal{H}^{\rm{SUSY}}=\mathcal{H}^{\rm{SUSY}}_{+}\oplus \mathcal{H}^{\rm{SUSY}}_-$ and $\Theta^+_{\pm\gamma}$ and $\Theta^-_{\pm\gamma}$ are grade-preserving super-operators. The Witten parity operator of the system $\mathcal{W}=\left(\begin{array}{cc}
\mathbf{1}_{2\times 2} & \mathbf{0}_{2\times 2} \\
\mathbf{0}_{2\times 2} & -\mathbf{1}_{2\times 2}
\end{array} \right)$ with the desired relations $[\mathcal{W}, \mathcal{H}^{\rm{SUSY}}]=\{\mathcal{W}, \Theta_{\pm\gamma}^\pm\}=\{\mathcal{W}, (\Theta_{\pm\gamma}^\pm)^{\dagger}\}=0$ and $\mathcal{W}^2=\mathbf{1}_{4\times 4}$.

The Hamiltonian $ H^{\rm{SUSY}}_{\pm\gamma}$ is also pseudo-supersymmetric in the sense of Mostafazadeh \cite{supermosta02} if one considers $\Delta^B_{\pm\gamma}=\wp^B_{\pm\gamma}$ and $(\Delta^B_{\pm\gamma})^{\#}=\mathcal{T}^{-1}(\wp^B_{\pm\gamma})^{\dagger}\mathcal{T}$  (pseudo-self-adjointness) and writes $H^{\rm{pSUSY}}_{\pm\gamma}=\{\Lambda^+_{\pm\gamma}, \Lambda^-_{\pm\gamma}\}$, where,

\begin{equation}
\Lambda^+_{\pm\gamma}=\frac{1}{\sqrt{2}}\left(\begin{array}{cc}
\mathbf{0}_{2\times 2} & \Delta^B_{\pm\gamma} \\
\mathbf{0}_{2\times 2} & \mathbf{0}_{2\times 2}
\end{array} \right)\:\:{\rm{and}}\:\:\Lambda^-_{\pm\gamma}=\frac{1}{\sqrt{2}}\left(\begin{array}{cc}
\mathbf{0}_{2\times 2} & \mathbf{0}_{2\times 2}\\
 (\Delta^B_{\pm\gamma})^{\#} & \mathbf{0}_{2\times 2}
\end{array} \right).
\end{equation}
 It is immediate to see $H^{\rm{SUSY}}_{\pm\gamma}=H^{\rm{pSUSY}}_{\pm\gamma}$, intertwining relations $R^+_{\pm\gamma} (\Delta^B_{\pm\gamma})= (\Delta^B_{\pm\gamma})R^-_{\pm\gamma}$ and $R^-_{\pm\gamma} (\Delta^B_{\pm\gamma})^{\#}= (\Delta^B_{\pm\gamma})^{\#}R^+_{\pm\gamma}$ along with $\mathfrak{T}^{-1}(\Lambda^+_{\pm\gamma})^{\dagger}\mathfrak{T}=(\Lambda^+_{\pm\gamma})^{\#}$, where $\mathfrak{T}=\left(\begin{array}{cc}
\mathcal{T} & \mathbf{0}_{2\times 2} \\
\mathbf{0}_{2\times 2} & \mathcal{T} 
\end{array} \right)$ may be called super-time-reversal operator. The pseudo-SUSY partners share mutually time reversed generators in their expressions and give back  SUSY partners for $\gamma=0$. Many other SUSY and pseudo-SUSY related results may be discussed elsewhere.

\section{Possible Physical Application of the Model}
The Rashba interaction term can be understood through the expression $\beta\mathbf{\Omega}\cdot\mathbf{e^\gamma}$, where the vector operator $\mathbf{\Omega}=P_2\mathbf{u}_1-P_1\mathbf{u}_2$ acts exactly like a magnetic field (reminiscent of Zeeman term or Pauli Hamiltonian) resulting to the phenomenon of spin precession. For a many electron system electrons with different momenta precess around different axes. Therefore, scattering involving different momenta randomizes the precession of a polarized ensemble and consequently leads to spin relaxation known as \textbf{D'yakonov Perel' Effect} \cite{long16, tuan16}. This mechanism along with many other (where the interactions are predominantly of spin-orbit type) have already been understood by means of momentum dependent density matrix \cite{long16} which is used in \textbf{spin-Boltzmann} and \textbf{spin continuity} equation \cite{fabi07}. In some occasions spin induced \textbf{Time Reversal Symmetry Breaking} and allied de-phasing have also been reported \cite{meijer05}. Our present forms of interaction term and spinors obtained from this model show that they can be manipulated by changing the value of the parameter $\gamma$, responsible for non-Hermiticity.  In fact, the orientations of the bi-orthogonal spinors are dependent on $\gamma$ as has been emphasized before (section-\ref{biortho}). In view of this, it is suggestive to investigate both theoretically and experimentally, spin relaxation mechanism and similar observations when the spin is manipulated (by changing $\gamma$) in place of momentum, a possibility that is not conceivable for the corresponding Hermitian model.

\section{Conclusion} Bi-orthogonal quantum formalism is not yet a well-settled issue especially in infinite dimension \cite{brody14}. Nevertheless, its importance is increasingly felt due to its close association with pseudo-Hermitian quantum mechanics and many ideas well accepted in Hermitian Quantum Mechanics need renewed attention in the bi-orthogonal formalism. The present article however does not attempt to review the unsettled issues. It rather deals with a case ($\mathcal{T}$-pseudo-Hermitian Rashba Hamiltonian) where the bi-orthogonal eigen-spinors involve a free-particle part and the question of bi-orthogonality is addressed in various ways in a typical Clifford algebraic frame-work where, spinors are viewed as left ideal of the the real Clifford algebra $Cl_3 (\mathbb{R})$. The fermionic time reversal operator paves the way to deal with the generators of $Cl_3(\mathbb{R})$ and their time-reversed counter parts. This requires an extensive use of various involutions understood in terms of conventional $Cl_3(\mathbb{R})$ generators. An analogous version of Kramers' theorem has been developed though conjectural in sense for the present Hamiltonian. Some preliminary  features of super-symmetry and pseudo-super-symmetry have been observed in the present context with the later being a manifestly new feature in the parlance of pseudo-Hermitian Quantum Mechanics. Several spin related phenomena related to many electron system may be investigated in view of this model. Similar approach to Dresselhaus spin-orbit interaction can be considered for further explore.

\section*{Acknowledgement} A.C. wishes to thank his colleague Dr. Baisakhi Mal for her valuable assistance in preparing the latex version.

\section*{ORCID iDs}
Arindam Chakraborty https://orcid.org/0000-0002-3414-3785


\section*{References}
\begin{thebibliography}{99}
\bibitem{benderbook} Bender C M 2019 {\it PT symmetry in Quantum and Classical Physics}   (London (UK) : World Scientific)
\bibitem{moise11} Moiseyev N 2011 {\it Non-Hermitian Quantum Mechanics}   (Chambridge(UK) : Chembridge University Press)
\bibitem{baga15} Albeverio S,Antoine J-P,Bagarello F,Caliceti E,Graffi S,Kuzhel S,KrejcirikD, Siegl P, Szafraniec
FH, Trapani C, and Znojil M 2015 Non-Selfadjoint Operators in Quantum Physics ed F Bagarello,
F H Szafraniec, M Znojil and J-P Gazeau (New York:Wiley) (ch 3 pp 123–124 and pp 277–278
regarding basis construction), (ch 6 specifically pp 323–324 regarding PT symmetry)
\bibitem{bender98} Bender C M and Boettcher S 1998 Real Spectra of Non-Hermitian Hamiltonian having $\mathcal{PT}$ symmetry {\it Phys. Rev. Lett.} $\bf{80}$  5243
\bibitem{bender02} Bender C M,  Brody D C and  Jones H F 2002 Complex Extension of Quantum Mechanics {\it Phys. Rev. Lett.} $\bf{89}$  27040
\bibitem{brody16} Brody D C 2016 Consistency of PT symmetric Quantum Mechanics {\it J. Phys. A: Math. Theor.} $\bf{49}$  10LT03.
\bibitem{mosta02} Mostafazadeh A 2002 Pseudo-Hermiticity versus PT symmetry : The necessary condition for the reality of the spectrum of a non-Hermitian Hamiltonian {\it J. Math. Phys.} $\bf{43}$ 205
\bibitem{mosta02a} Mostafazadeh A 2002 Pseudo-Hermiticity versus PT symmetry. II.  A complete characterization of non-Hermitian Hamiltonians with real spectrum {\it J. Math. Phys.} $\bf{43}$ 2814
\bibitem{mosta02b} Mostafazadeh A 2002 Pseudo-Hermiticity for a class of nondiagonalizable Hamiltonians {\it J. Math. Phys.} $\bf{43}$ 6343
\bibitem{znojil04} Znojil M  2004 Fragile PT symmetry in a solvable model {\it J. Math. Phys.} $\bf{45}$ 4418
\bibitem{beygi15} Beygi A,  Klevansky S P and  Bender C M 2015 {\it Phys. Rev. A.} $\bf{91}$ 062101
\bibitem{chakraborty20} Chakraborty A 2020 A new boson realization of fusion polynomial algebras in non-Hermitian
quantum mechanics: $\gamma$-deformed su(2) generators, $\mathcal{PT}$ -symmetry in Fock space and Higgs algebra {\it J. Phys. A: Math. Theor.} $\bf{53}$ 485202
\bibitem{chakraborty21} Chakraborty A 2021 Understanding Partial $\mathcal{PT}$ Symmetry as Weighted Composition Conjugation in Reproducing Kernel Hilbert Space: An application to Non-hermitian Bose-Hubbard Type Hamiltonian in Fock space {\it Int. J. Theor. Phys.} $\bf{60}$ 3689
\bibitem{brody14} Brody D C 2014 Biorthogonal quantum mechanics {\it J. Phys. A: Math. Theor.} $\bf{47}$ 035305
\bibitem{curt07} Curtright T, Mezincescu L 2007 Biorthogonal quantum systems {\it J. Math. Phys.} $\bf{48}$ 092106
\bibitem{curt07a} Curtright T, Mezincescu L, Schuster D 2007 Supersymmetric biorthogonal quantum systems {\it J. Math. Phys.} $\bf{48}$ 092108
\bibitem{houn17} Hounguevou J V,  Dossa F A,  Avossevou G Y H 2017  Biorthogonal Quantum Mechanics For Non-Hermitian Multimode and Multiphoton Jaynes–Cummings Models {\it Theoretical and Mathematical Physics} $\bf{193}$ 1464
\bibitem{oscar18} Oscar R O, Kevin Z 2018 Bi-orthogonal approach to non-Hermitian Hamiltonians with the oscillator spectrum: Generalized coherent states for nonlinear algebras {\it Annals of Physics.} $\bf{388}$ 26
\bibitem{chang13} Chang L M, Lewis Z, Minic D, Takeuchi T 2013 Biorthogonal quantum mechanics: super-quantum
correlations and expectation values without definite probabilities {\it J. Phys. A: Math. Theor.} $\bf{46}$ 485306
\bibitem{Ra60}  Sch\"apers T  2016 {\it Semiconductor Spintronics.}  (Berlin/Boston : Walter de Gruyter) ch 7 pp 152
\bibitem{Lu10}  Lu Jian-Duo 2010  { Electron tunneling in a non-magnetic heterostructure in presence of both Dresselhaus and Rashba spin-orbit terms. } {\it Physica E : Low-dimensional Systems and Nanostucture} \textbf{43} 142
\bibitem{YG14} Yan Xu, Gu Qiang 2014 { Superconductivity in a two-dimensional superconductor with Rashba and Dresselhaus spin-orbit couplings. } {\it Solid State Communication} \textbf{187}  68.
\bibitem{masa09} Sato M,  Fujimoto S 2009 {Topological phases of noncentrosymmetric superconductors: Edge states, Majorana fermions, and non-Abelian statistics.} {\it Phys. Rev. B} \textbf{79}  094504
\bibitem{zhang05}  Zhang Xi-Hua,  Xiong Shi-Jie 2005 Scaling behavior of transport properties in one-dimensional
network with both Rashba spin–orbit coupling and disorder {\it Physica B} $\bf{368}$ 261
\bibitem{zhang08}  Zhang Hai-Rui,  Xiao Jing-Lin 2008 The effective mass of strong-coupling polaron in a triangular quantum well induced by the Rashba effect {\it Physica B} $\bf{403}$ 1933
\bibitem{rashba88} Rashba E I and  Sherman E Ya 1988 Spin-Orbital band splitting in symmetry quantum wells {\it Phys. Lett. A} \textbf{129} 175
\bibitem{xu08} Xu Qing-Qiang, Gao  Ben-Ling,  Xiong Shi-Jie 2008 Flux-dependent Kondo temperature with local Rashba and Coulomb interactions in the Kondo regime {\it Physica B} $\bf{403}$ 1686
\bibitem{byrashba84}  Bychkov Yu A,  Rashba E I 1984 { Oscillatory effects and the magnetic susceptibility of carriers in inversion layers. } {\it J. Phys C: Solid State Physics} $\bf{17}$  6039.
\bibitem{kozin18}  Kozin V  K, Iorsh I V,  Kibis O V, and  Shelykh I A 2018 Quantum ring with the Rashba spin-orbit interaction in the regime of strong light-matter coupling {\it Phys. Rev. B} \textbf{97} 155434
\bibitem{jones10}  Jones-Smith K and  Mathur H 2010 Non-Hermitian quantum Hamiltonians with $\mathcal{PT}$ symmetry{\it Phys. Rev. A} \textbf{82} 042101
\bibitem{bender11}  Bender C M, Klevansky S P 2011 $\mathcal{PT}$-symmetric representations of fermionic algebras {\it Phys. Rev. A} \textbf{84} 024102
\bibitem{jones14}  Jones-Smith K and  Mathur H 2014 Relativistic non-Hermitian quantum mechanics {\it Phys. Rev. D} \textbf{89} 125014
\bibitem{choutri14}  Choutri B, Cherbal O,  Ighezou F Z, and Trifonov D A 2014 On the time-reversal symmetry in
pseudo-Hermitian systems  {\it Prog. Theor. Exp. Phys. }   113A02
\bibitem{lahiri88}  Lahiri A,  Roy P K, Bagchi B 1988 Isospectral Hamiltonians for potentials having rotational symmetry in N dimensions {\it Phys. Rev. A}  \textbf{38} 6419 
\bibitem{sinha02}  Sinha A,  Roychoudhury R 2002 Isospectral partners of a complex $\mathcal{PT}$-invariant potential {\it Phys. Lett. A}  \textbf{301} 163 
\bibitem{sanjay96} Sanjay Kumar M, Khare A 1996 Coherent states for isospectral Hamiltonians {\it Phys. Lett. A} \textbf{217} 73
\bibitem{naru97}  Aizawa N,  Sato Haru-Tada 1997 Isospectral Hamiltonians and $W_{1+\infty}$ Algebra {\it Prog. Theor. Phys.} \textbf{98} 707
\bibitem{pursey86}  Pursey D L 1986 Isometric operators, isospectral Hamiltonians, and supersymmetric quantum mechanics {\it Phys. Rev. D} \textbf{33} 2267
\bibitem{pursey86a}  Pursey D L 1986 New families of isospectral Hamiltonians {\it Phys. Rev. D} \textbf{33} 1048
\bibitem{gomes02}  Lima V G, Santos V S,  Rodrigues R. de Lima 2002 On the scalar potential models from the isospectral potential class {\it Phys. Lett. A} \textbf{298} 91
\bibitem{sid17} Yahiaoui Sid-Ahmed  and  Bentaibaa M 2017 Isospectral Hamiltonian for position dependent mass for an arbitrary quantum system and coherent states {\it J. Math. Phys.} $\bf{58}$ 063507
\bibitem{geru18} Geru, I I 2018 {\it Time-Reversal Symmetry : Seven Time Reversal Operators for Spin Containing System} (Switzerland AG : Springer Nature) ch 2 pp 40-54
\bibitem{klein51}  Klein M J 1952 On a Degeneracy Theorem of Kramers {\it Am. J. Phys.} \textbf{20} 65
\bibitem{scharf87}  Scharf R, Dietz B, K\'us M, Haake F, Berry M V 1988 Kramers' Degeneracy and Quartic Level Repulsion {\it Europhys. Lett.} \textbf{5} 383
\bibitem{kruthoff19}   Kruthoff J,  de Boer J,  van Wezel J 2019 Topology in time-reversal symmetric crystals {\it Phys. Rev. B} \textbf{100} 075116
\bibitem{rosch83}  R\"osch N 1983 Time-reversal Symmetry, Kramers' Degeneracy and the Algebraic Eigenvalue Problem {\it Chem. Phys.} \textbf{80} 1 
\bibitem{konstantin20}  Pichugin K,  Puente A,   Nazmitdinov R 2020 Kramers Degeneracy and Spin Inversion in a Lateral
Quantum Dot {\it Symmetry} \textbf{12} 2043 
\bibitem{revaz09} Ramazashvili R 2009 Kramers degeneracy in a magnetic field and Zeeman spin-orbit coupling
in antiferromagnetic conductors {\it Phys. Rev. B} \textbf{79} 184432
\bibitem{chen22}  Zhang P,  Chen Y 2022 Violation and revival of Kramers’ degeneracy in open quantum systems {\it Phys. Rev. B} \textbf{105} L241106
\bibitem{lieu22}  Lieu S,  McGinley M,  Shtanko O,  Cooper N R, and  Gorshkov A V 2022 Kramers' degeneracy for open systems in thermal equilibrium {\it Phys. Rev. B} \textbf{105} L121104
\bibitem{junker19} Junker G 2019 {\it Supersymmetric Methods in Quantum, Statistical and Solid State Physics}  (IOP Publishing (UK))
\bibitem{supermosta01}  Samani K A, Mostafazadeh A 2001 {Quantum Mechanical Symmetries and Topological Invariants. } 2012 {\it Nucl. Phys B}  \textbf{595}  467.
\bibitem{supermosta02}  Mostafazadeh A 2002 {Pseudo-supersymmetric quantum mechanics and isospectral pseudo-Hermitian Hamiltonians}  {\it Nucl. Phys B}  \textbf{640 [PM]}  419.
\bibitem{alex20}  Alexandre J,  Ellis J , Millington P 2020 {PT -symmetric non-Hermitian quantum field theories with supersymmetry} {\it Phys. Rev. D} \textbf{101} 085015.
\bibitem{superfring20}  Cen J,  Fring A,  Frith T 2020 {Time-dependent Darboux (supersymmetric)
transformations for non-Hermitian quantum systems} {\it J. Phys. A : Math. and Theor.} \textbf{52} 115302.
\bibitem{andri07}  Andrianov A A,  Cannatab F,  Sokolov A V 2007 { Non-linear supersymmetry for non-Hermitian,
non-diagonalizable Hamiltonians: I. General properties} {\it Nucl. Phys. B } \textbf{773 [PM]} 107.
\bibitem{superznojil02} Znojil M 2002 {Non-Hermitian supersymmetry and singular, $\mathcal{PT} $-symmetrized oscillators} {\it J. Phys. A : Math. Theor.} \textbf{35} 2341.
\bibitem{wang17} Wang L, 2017 {Superconductivity in a two-dimensional repulsive Rashba gas at low electron density} {\it J. Phys. Commun.} \textbf{1} 011001
\bibitem{yes11} Yesiltas  \"O 2011 {Quantum isotonic nonlinear oscillator as a Hermitian counterpart of Swanson Hamiltonian and pseudo-supersymmetry} {\it J. Phys. A : Math. and Theor.} \textbf{44} 305305.
\bibitem{sinha08}  Sinha A,  Roy P 2008 {Pseudo supersymmetric partners for the generalized Swanson model} {\it J. Phys. A : Math. and Theor.} \textbf{41} 335306.
\bibitem{bagchi15} Bagchi B,  Banerjee A, Mandal P 2015 {A generalized Swanson Hamiltonian in a second-derivative pseudo-supersymmetric framework} {\it Int. J. Mod. Phys. A} \textbf{30} 1550037.
\bibitem{LL67}  L\'evy-Leblond J M 1967 { Nonrelativistic Particles and Wave Equations.} {\it Commun. Math. Phys.} \textbf{6} 286.
\bibitem{Hl99} Hladik J 1999 {\it Spinors in Physics.} New York : Springer Science+Business media, LLC 
\bibitem{monti06}  de Montigny M,  Niederle J,  Nikitin A G 2006 { Galilei invariant theories: I. Constructions of indecomposable finite-dimensional representations of the homogeneous Galilei group: directly and via contractions} {\it J. Phys. A Math. Gen.} \textbf{39}  9365.
\bibitem{nieder09}  Niederle J,  Nikitin A G 2009 { Galilean equations for massless fields.} {\it J. Phys. A :  Math. Theor.} \textbf{42}  105207.
\bibitem{hue12}  Huegele R,  Musielak Z E,  Fry J L, { Fundamental dynamical equations for spinor wavefunctions: I. Lévy-Leblond and Schrödinger equations. } 2012 {\it J. Phys. A : Math.Theor.}  \textbf{45}  145205.
\bibitem{hue13}  Huegele R,  Musielak Z E,  Fry J L 2013 { Generalized L\'evi-Leblond and Schr\"odinger  Equations
for Spinor Wavefunctions.} {\it Adv. Studies in Theor. Phys.} \textbf{7}  825.
\bibitem{deri10} Deriglazov A, 2010 {\it Classical Mechanics (Hamiltonian and Lagrangian Formalism)}
 (Berlin : Springer) ch 1 pp 54
\bibitem{lanczos70} Lanczos C, 1986 {\it The Variational Principles of Mechanics.} Unabridged Dover (1986) Republication of the fourth edition published by University of Toronto Press, Toronto (1970) ch V point 5
\bibitem{wu05}  Xu W,  Guo Y 2005 {Rashba and Dresselhaus spin-orbit coupling effects on tunnelling through two-dimensional magnetic quantum systems} {\it Phys. Lett. A} \textbf{340} 281.
\bibitem{vasil06} Vasilopoulos P,  Wang X F 2006 {Spin-dependent magnetotransport through one or more rings
in the presence of the Rashba and Dresselhaus terms of the spin-orbit interaction} {\it Physica E} \textbf{34} 359
\bibitem{winkler03}  Winkler R 2003 {\it Spin-Orbit Coupling Effects in Two-Dimensional Electron and Hole Systems.} (Berlin, Heidelberg (Germany) : Spinger-Verlag) chap 5, 6
\bibitem{Lo01}  Lounesto P 2001 {\it Clifford Algebras and Spinors.} ( Cambridge (UK) : Cham. Univ. Press ) ch 1, 2, 3, 4
\bibitem{vaz16} Vaz Jr. J and  Rocha Jr. R D 2016 {\it An Introduction to Clifford Algebras and Spinors.} (Oxford (UK) : Oxford. Univ. ) ch 1, 2, 4
\bibitem{chakraborty22} Chakraborty A, Debnath B, Dutta R, Banerjee P 2022 Construction of a few Quantum mechanical Hamiltonians via L\'evy-Leblond type Linearization: Clifford momentum, Spinor states and Supersymmetry {\it Adv. Appl. Clifford Algebras} $\bf{32}$ 56
\bibitem{gelfand64} Gel'fand I M, Vilenkin N Ya, 1964 {\it Generalized Functions vol-4  Applications of Harmonic Analysis}  (Academic Press, New York, London) (chap-4 pp 103)
\bibitem{wang04}  Wang  J,  Sun H B and Xing D Y 2004 { Rashba spin precession in a magnetic field} {\it Phys. Rev. B} \textbf{69} 085304
\bibitem{aver02}  Averkiev N S,  Golub L E  and  Willander M 2002 Spin relaxation anisotropy in two-dimensional
semiconductor systems {\it  J Phys Condens. Matter} \textbf{14} R271
\bibitem{long16}  Long N H,  Mavropoulos P,  Bauer D. S. G,  Zimmermann B,  Mokrousov Y, and  Bl\"ugel S  2016 { Strong spin-orbit fields and Dyakonov-Perel spin dephasing in supported metallic films} {\it Phys Rev.  B} \textbf{94} 180406(R)
\bibitem{tuan16}  Tuan D V,  Adam S and  Roche S 2016 {Spin dynamics in bilayer graphene: Role of electron-hole puddles and Dyakonov-Perel mechanism} {\it Phys Rev B} \textbf{94}  041405(R)
\bibitem{hajek08} H$\acute{\rm a}$jek P,  Santaluc$\acute{\rm i}a$ V M,  Vanderwerff J,  Zizler V 2008 {\it Bi-Orthogonal Systems in Banach Spaces}  (New York : Springer, Science+Bussiness Media, LLC ) (ch 1 pp 5 theorem-1.16)
\bibitem{meijer05}  Meijer F E, Morpurgo A F,  Klapwijk T  M  and  Nitta J 2005 Universal Spin-Induced Time Reversal Symmetry Breaking in Two-Dimensional Electron Gases with Rashba Spin-Orbit Interaction {\it Phys. Rev. Lett.} \textbf{94} 186805
\bibitem{fabi07}  Fabian J,  Matos-Abiaguea A,  Ertlera C,  Stano P,   ${\rm{\breve{Z}}}$utic I 2007 Semiconductor  Spintrocics {\it acta physica slovaca} \textbf{57} 565 (pp-720)
\end{thebibliography}
\end{document}